\RequirePackage[left,columnwise]{lineno}
\documentclass[aps,prc,twocolumn,floatfix,superscriptaddress,nofootinbib]{revtex4}

\usepackage{lipsum}
\usepackage{graphicx}
\usepackage{epstopdf}
\usepackage{subfigure}
\usepackage{epsfig}
\usepackage{amsmath,amssymb,amsfonts,eufrak,mathrsfs}
\usepackage{bm}
\usepackage{color}
\usepackage{lineno}

\newcommand{\bea}{\begin{eqnarray}}
\newcommand{\eea}{\end{eqnarray}}

\usepackage[utf8]{inputenc}

\begin{document}


\title{Hybrid Hadronization - A Study of In-Medium Hadronization of Jets}


\author{A.~Sengupta}
\email{aseng@tamu.edu}
\affiliation{Cyclotron Institute, Texas A\&M University, College Station TX 77843.}
\affiliation{Department of Physics and Astronomy, Texas A\&M University, College Station TX 77843.}

\author{R.~J.~Fries}
\email{rjfries@comp.tamu.edu}
\affiliation{Cyclotron Institute, Texas A\&M University, College Station TX 77843.}
\affiliation{Department of Physics and Astronomy, Texas A\&M University, College Station TX 77843.}

\author{M.~Kordell II}
\email{mkordell@tamu.edu}
\affiliation{Cyclotron Institute, Texas A\&M University, College Station TX 77843.}
\affiliation{Department of Physics and Astronomy, Texas A\&M University, College Station TX 77843.}

\author{B.~Kim}
\affiliation{Cyclotron Institute, Texas A\&M University, College Station TX 77843.}
\affiliation{Department of Physics and Astronomy, Texas A\&M University, College Station TX 77843.}





\author{A.~Angerami}
\affiliation{Lawrence Livermore National Laboratory, Livermore CA 94550.}

\author{R.~Arora}
\affiliation{Department of Computer Science, Wayne State University, Detroit MI 48202.}

\author{S.~A.~Bass}
\affiliation{Department of Physics, Duke University, Durham, NC 27708.}

\author{Y.~Chen}
\affiliation{Laboratory for Nuclear Science, Massachusetts Institute of Technology, Cambridge MA 02139.}
\affiliation{Department of Physics, Massachusetts Institute of Technology, Cambridge MA 02139.}
\affiliation{Department of Physics and Astronomy, Vanderbilt University, Nashville TN 37235.}

\author{R.~Datta}
\affiliation{Department of Physics and Astronomy, Wayne State University, Detroit MI 48201.}

\author{L.~Du}
\affiliation{Department of Physics, McGill University, Montr\'{e}al QC H3A\,2T8, Canada.}
\affiliation{Department of Physics, University of California, Berkeley CA 94270.}
\affiliation{Nuclear Science Division, Lawrence Berkeley National Laboratory, Berkeley CA 94270.}

\author{R.~Ehlers}
\affiliation{Department of Physics, University of California, Berkeley CA 94270.}
\affiliation{Nuclear Science Division, Lawrence Berkeley National Laboratory, Berkeley CA 94270.}

\author{H.~Elfner}
\affiliation{GSI Helmholtzzentrum f\"{u}r Schwerionenforschung, 64291 Darmstadt, Germany.}
\affiliation{Institute for Theoretical Physics, Goethe University, 60438 Frankfurt am Main, Germany.}
\affiliation{Frankfurt Institute for Advanced Studies, 60438 Frankfurt am Main, Germany.}

\author{C.~Gale}
\affiliation{Department of Physics, McGill University, Montr\'{e}al QC H3A\,2T8, Canada.}

\author{Y.~He}
\affiliation{School of Physics and Optoelectronics, South China University of Technology, Guangzhou 510640, China}

\author{B.~V.~Jacak}
\affiliation{Department of Physics, University of California, Berkeley CA 94270.}
\affiliation{Nuclear Science Division, Lawrence Berkeley National Laboratory, Berkeley CA 94270.}

\author{P.~M.~Jacobs}
\affiliation{Department of Physics, University of California, Berkeley CA 94270.}
\affiliation{Nuclear Science Division, Lawrence Berkeley National Laboratory, Berkeley CA 94270.}

\author{S.~Jeon}
\affiliation{Department of Physics, McGill University, Montr\'{e}al QC H3A\,2T8, Canada.}

\author{Y.~Ji}
\affiliation{Department of Statistical Science, Duke University, Durham NC 27708.}

\author{F.~Jonas}
\affiliation{Department of Physics, University of California, Berkeley CA 94270.}
\affiliation{Nuclear Science Division, Lawrence Berkeley National Laboratory, Berkeley CA 94270.}

\author{L.~Kasper}
\affiliation{Department of Physics and Astronomy, Vanderbilt University, Nashville TN 37235.}

\author{A.~Kumar}
\affiliation{Department of Physics, University of Regina, Regina, SK S4S 0A2, Canada.}
\affiliation{Department of Physics, McGill University, Montr\'{e}al QC H3A\,2T8, Canada.}

\author{R.~Kunnawalkam-Elayavalli}
\affiliation{Department of Physics and Astronomy, Vanderbilt University, Nashville TN 37235.}

\author{J.~Latessa}
\affiliation{Department of Computer Science, Wayne State University, Detroit MI 48202.}

\author{Y.-J.~Lee}
\affiliation{Laboratory for Nuclear Science, Massachusetts Institute of Technology, Cambridge MA 02139.}
\affiliation{Department of Physics, Massachusetts Institute of Technology, Cambridge MA 02139.}

\author{R.~Lemmon}
\affiliation{Daresbury Laboratory, Daresbury, Warrington, Cheshire, WA44AD, United Kingdom.}

\author{M.~Luzum}
\affiliation{Instituto  de  F\`{i}sica,  Universidade  de  S\~{a}o  Paulo,  C.P.  66318,  05315-970  S\~{a}o  Paulo,  SP,  Brazil. }

\author{A.~Majumder}
\affiliation{Department of Physics and Astronomy, Wayne State University, Detroit MI 48201.}

\author{S.~Mak}
\affiliation{Department of Statistical Science, Duke University, Durham NC 27708.}

\author{A.~Mankolli}
\affiliation{Department of Physics and Astronomy, Vanderbilt University, Nashville TN 37235.}

\author{C.~Martin}
\affiliation{Department of Physics and Astronomy, University of Tennessee, Knoxville TN 37996.}

\author{H.~Mehryar}
\affiliation{Department of Computer Science, Wayne State University, Detroit MI 48202.}

\author{T.~Mengel}
\affiliation{Department of Physics and Astronomy, University of Tennessee, Knoxville TN 37996.}

\author{C.~Nattrass}
\affiliation{Department of Physics and Astronomy, University of Tennessee, Knoxville TN 37996.}

\author{J.~Norman}
\affiliation{Oliver Lodge Laboratory, University of Liverpool, Liverpool, United Kingdom.}

\author{C.~Parker}
\affiliation{Cyclotron Institute, Texas A\&M University, College Station TX 77843.}
\affiliation{Department of Physics and Astronomy, Texas A\&M University, College Station TX 77843.}

\author{J.-F.~Paquet}
\affiliation{Department of Physics and Astronomy, Vanderbilt University, Nashville TN 37235.}

\author{J.~H.~Putschke}
\affiliation{Department of Physics and Astronomy, Wayne State University, Detroit MI 48201.}

\author{H.~Roch}
\affiliation{Department of Physics and Astronomy, Wayne State University, Detroit MI 48201.}

\author{G.~Roland}
\affiliation{Laboratory for Nuclear Science, Massachusetts Institute of Technology, Cambridge MA 02139.}
\affiliation{Department of Physics, Massachusetts Institute of Technology, Cambridge MA 02139.}

\author{B.~Schenke}
\affiliation{Physics Department, Brookhaven National Laboratory, Upton NY 11973.}

\author{L.~Schwiebert}
\affiliation{Department of Computer Science, Wayne State University, Detroit MI 48202.}

\author{C.~Shen}
\affiliation{Department of Physics and Astronomy, Wayne State University, Detroit MI 48201.}
\affiliation{RIKEN BNL Research Center, Brookhaven National Laboratory, Upton NY 11973.}

\author{M.~Singh}
\affiliation{Department of Physics and Astronomy, Vanderbilt University, Nashville TN 37235.}

\author{C.~Sirimanna}
\affiliation{Department of Physics, Duke University, Durham, NC 27708.}
\affiliation{Department of Physics and Astronomy, Wayne State University, Detroit MI 48201.}

\author{D.~Soeder}
\affiliation{Department of Physics, Duke University, Durham, NC 27708.}

\author{R.~A.~Soltz}
\affiliation{Department of Physics and Astronomy, Wayne State University, Detroit MI 48201.}
\affiliation{Lawrence Livermore National Laboratory, Livermore CA 94550.}

\author{I.~Soudi}
\affiliation{Department of Physics and Astronomy, Wayne State University, Detroit MI 48201.}
\affiliation{University of Jyväskylä, Department of Physics, P.O. Box 35, FI-40014 University of Jyväskylä, Finland.}
\affiliation{Helsinki Institute of Physics, P.O. Box 64, FI-00014 University of Helsinki, Finland.}

\author{Y.~Tachibana}
\affiliation{Akita International University, Yuwa, Akita-city 010-1292, Japan.}

\author{J.~Velkovska}
\affiliation{Department of Physics and Astronomy, Vanderbilt University, Nashville TN 37235.}

\author{G.~Vujanovic}
\affiliation{Department of Physics, University of Regina, Regina, SK S4S 0A2, Canada.}

\author{X.-N.~Wang}
\affiliation{Key Laboratory of Quark and Lepton Physics (MOE) and Institute of Particle Physics, Central China Normal University, Wuhan 430079, China.}
\affiliation{Department of Physics, University of California, Berkeley CA 94270.}
\affiliation{Nuclear Science Division, Lawrence Berkeley National Laboratory, Berkeley CA 94270.}

\author{X.~Wu}
\affiliation{Department of Physics, McGill University, Montr\'{e}al QC H3A\,2T8, Canada.}
\affiliation{Department of Physics and Astronomy, Wayne State University, Detroit MI 48201.}

\author{W.~Zhao}
\affiliation{Department of Physics and Astronomy, Wayne State University, Detroit MI 48201.}
\affiliation{Department of Physics, University of California, Berkeley CA 94270.}
\affiliation{Nuclear Science Division, Lawrence Berkeley National Laboratory, Berkeley CA 94270.}

\collaboration{The JETSCAPE Collaboration}

\date{\today}


\begin{abstract}
QCD jets are considered important probes for quark gluon plasma created in collisions of nuclei at high energies. Their parton showers are significantly altered if they develop inside of a deconfined medium. Hadronization of jets is also thought to be affected by the presence of quarks and gluons.
We present a systematic study of the effects of a thermal bath of partons on the hadronization of parton showers. 
We use the JETSCAPE framework to create parton showers both in vacuum and in a brick of quark gluon plasma. The brick setup allows important parameters, like the size of the plasma as well as the collective flow of partons, to be varied systematically.
We hadronize the parton showers using Hybrid Hadronization, which permits shower partons to form strings with thermal partons, or to recombine directly with thermal partons as well as with each other. We find a sizeable amount of interaction of shower partons with thermal partons during hadronization, indicating a natural continuation of the interaction of jet and medium during this stage. The observed effects grow with the size of the medium. Collective flow easily transfers from the thermal partons onto the emerging jet hadrons. We also see a significant change in hadron chemistry as expected in the presence of quark recombination processes.
\end{abstract}

\maketitle


\section{Introduction}
\label{sec:introduction}

Hadronization has been a largely intractable problem since the inception of quantum chromodynamics (QCD) as the theory describing the strong nuclear force. The formation of bound states of quarks and gluons typically involves low momentum scales, placing them firmly in the non-perturbative region of QCD. 
However, simulations of perturbative QCD processes in Monte Carlo generators like PYTHIA \cite{Sjostrand:2006za,Sjostrand:2014zea}, HERWIG \cite{Marchesini:1991ch,Bellm:2015jjp} or JETSCAPE \cite{Putschke:2019yrg}, require a practical solution to this problem that can turn parton output into hadrons while capturing as many physical features of the hadronization process as possible. Thus the challenge is to construct comprehensive and physically motivated hadronization models. They should adhere to a few key principles: conservation of energy, momentum, and other quantum numbers; enforcement of confinement; consistency with color flow in the parton system;
and a reasonable hierarchy in the production of excited and ground state hadrons. 

Cluster hadronization \cite{Webber:1983if} and Lund string fragmentation \cite{Andersson:1983ia} are two well-established models which have been used successfully in various Monte Carlo simulations for systems like $e^+$+$e^-$ and $p$+$p$. Yet another model that has been around since the early days of QCD \cite{Das:1977cp}, is quark recombination.
It has found new relevance with the advent of data from nucleus-nucleus ($A$+$A$) collisions at the Relativistic Heavy Ion Collider (RHIC) in the 2000s \cite{Fries:2003kq,Fries:2004ej,Greco:2003xt,Greco:2003mm,Fries:2008hs}. 
Quark recombination can explain features in A$+$A collisions that were difficult to accommodate for other models, e.g.\ large baryon over meson ratios $\sim\mathcal{O}(1)$, and the scaling of elliptic flow $v_2$ of identified hadrons with the constituent quark number of the hadron \cite{Fries:2008hs}. The success regarding these particular observables comes from two key features of quark recombination. By using partons that exist prior to the hadronization process, the creation of (anti)baryons with three valence (anti)quarks is not penalized compared to the creation of mesons with two valence quarks. This is typically the case in other models due to the higher mass of the baryon, or the large mass of a diquark-antidiquark pair that needs to be created. In $A$+$A$ collisions, a second key feature in recombination models comes into play. They allow partons from hard processes to recombine with the "background" of soft or thermal partons in such collisions. One should interpret this as a natural continuation of the interactions of jet partons with partons from the background medium into the non-perturbative regime. Roughly speaking, at the end of their perturbative evolution through a parton medium jet partons continue to interact with medium partons but now rather form hadrons in the process. This is a particularly effective way to impart key properties of the thermalized parton phase onto semi-hard hadrons at momenta of a few GeV$/c$, especially the collective flow of partons
\cite{Fries:2008hs}.

Hybrid Hadronization is a hadronization model 
which combines a Monte Carlo implementation of quark recombination with string fragmentation \cite{Han:2016uhh}. It is offered as a hadronization module in the JETSCAPE and XSCAPE frameworks \cite{Putschke:2019yrg}. The hybrid model was created to establish a unified, comprehensive description of hadronization that can be used in systems as diverse as $e^+$+$e^-$ and $A$+$A$, while at the same time utilizing the traditional strengths of both models in their respective areas.
The cutoff between the two components in the model is naturally determined particle-by-particle as recombination probabilities die off at large distances between partons in phase space. Roughly speaking, partons at small distances in phase space will prefer to hadronize by direct recombination into hadrons, while partons at large distances from other partons will have recombination probabilities which are small. These latter partons therefore have to connect to others by strings which eventually fragment into hadrons as well. 

Hybrid Hadronization was developed with 
applications to parton showers
in mind. It is therefore suitable for applications in vacuum systems like
$e^+$+$e^-$ and $p$+$p$ collisions. From recombination the model inherits the ability to handle an embedding of  shower partons into a thermal or non-thermal background medium. "Shower-thermal"
recombination processes can occur in this model as well as strings with built-in thermal partons. It is the application of Hybrid Hadroniztion to jets embedded in a background medium which is the main focus of this work.

The goal of this paper is two-fold. First, we want to demonstrate that Hybrid Hadronization can indeed lead to significant interactions between showers and medium partons at hadronization. We will show that these effects switch on smoothly if a medium is added, and that they grow with the size of the medium. This is a necessary condition that Hybrid Hadronization needs to pass in order to be a comprehensive model for all types of systems. Second, we will systematically study the effects of these interactions on common observables, like spectra, identified hadron ratios, and jet shapes. We will in particular study how collective flow of the background medium, either longitudinal or transverse to the jet, is imprinted on hadrons correlated with the jet. Note that direct comparison to data, or "tuning" of the hadronization model is not the goal here. A separate effort is ongoing to describe experimental data in vacuum systems like $e^+$+$e^-$ and $p$+$p$ collisions. Both efforts together will eventually be able to inspire the confidence to take a comprehensive look at data in $p$+$A$ and $A$+$A$ collisions. 

We expect that signature features of recombination, in particular enhanced baryon-over-meson ratios, and collective flow effects, can be reproduced well by Hybrid Hadronization. Our setup will allow us to study the dependences of the observables on parameters, like the size of the QGP medium, and the size and direction of the collective flow of the medium. We will focus here on jets with light flavors --- up, down and strange --- and we leave medium effects on hadronizing heavy quarks to a future discussion.

Tor this systematic study we use the JETSCAPE framework to simulate partons which are generated with a set initial energy and traverse a brick of quark gluon plasma. The medium partons inside the brick are kept at a uniform temperature and flow field, which can be varied. MATTER \cite{Majumder:2013re} is used to generate a shower from the initial parton from some initial virtuality. In-medium effects in MATTER are switched
on if a parton of the shower is inside the brick. Low-virtuality partons which are still inside the brick are subsequently handed over to LBT \cite{Cao:2016gvr} which simulates their further propagation through the brick. A parton is deemed ready to hadronize if
its virtuality is below a certain threshold $Q_0$, and it is located outside the brick, where there is no ambient medium.  Hadronization and decays of excited hadrons are then carried out using Hybrid Hadronization and calls to PYTHIA 8.

The paper is organized as follows. In the next section we describe key features of the Hybrid Hadronization model and its implementation in JETSCAPE. In Sec.\ \ref{sec:js_setup} we briefly lay out the setup in JETSCAPE used in this study, including important parameters. Sec. \ref{sec:results} discusses our results. We conclude with a summary and outlook.

\section{Hybrid Hadronization} 
\label{sec:hybhad}

In this section we briefly discuss some key aspects of Hybrid Hadronization with a particular focus on the implementation in JETSCAPE 3.0. Hybrid Hadronization expects two types of input, a list of jet shower partons, and, as an option, information on a thermal background medium. The following algorithm ensures that all jet shower partons  end up in hadrons, either by recombination, or string fragmentation. Thus confinement is enforced for shower partons. In contrast, thermal background partons are not mandated to hadronize but rather serve as a reservoir with which shower partons can interact. Excluding the creation of hadrons solely from the background is by choice, assuming better methods are available to create hadrons from thermal partons, e.g.\ by particlization in a fluid dynamic description of the medium \cite{Huovinen:2012is}. 
Hybrid Hadronization
samples thermal partons from the $T=T_c$ hypersurface of a fluid dynamic simulation that can be specified. For the current work the geometry is much simpler as the QCD medium is taken to be a brick of quark gluon plasma with a certain temporal size $L/c$. The hadronization hypersurface is thus determined by $t=L/c$  and partons are sampled for $T=T_c$ in this simple setup. 

Hybrid Hadronization requires space-time information to be read in with each parton in the parton shower. This is needed as recombination probabilities are evaluated in phase space. In this formalism phase space coordinates given for partons are assumed to be the centroids of Gaussian wave packets \cite{Han:2016uhh}. Space-time information is often not tracked in shower Monte Carlo generators that are not primarily aimed at nucleus-nucleus collisions. However, for A$+$A collisions space-time information is crucial. Both MATTER and LBT \cite{Cao:2016gvr} in JETSCAPE provide sufficient space-time information.

Hybrid Hadronization utilizes color tags to track color flow among shower partons. Color tags are provided by both PYTHIA 8 and MATTER in the vacuum. Showers going through a medium usually
have randomized color flow as the medium acts as a reservoir 
for color. Some partons, e.g.\ from LBT, have color tags set
to zero which indicates a random color.
Likewise, all partons in the list of sampled thermal partons 
are assigned color tags zero to indicate randomization. 

With all required input specified, Hybrid Hadronization provisionally forces gluons to decay into quark-antiquark pairs. Provisionally here means that the decay is not finalized if neither the quark nor antiquark daughter from the decay are used during the subsequent recombination stage. In that case, the gluon is put back into the parton list at the end of the recombination stage. The resulting list of shower and thermal quarks is randomized to avoid any bias, and handed over to the recombination module.

\subsection{Quark Recombination}
\label{subsec:reco}

The recombination stage of Hybrid Hadronization has been discussed in \cite{Han:2016uhh}. An accounting of the underlying formalism to compute probabilities for coalescence
of quarks in phase space can be found in \cite{Kordell:2021prk}. In a nutshell, Wigner distributions of mesons and baryons are computed assuming a 3-D isotropic harmonic oscillator potential between quarks. The characteristic sizes of the potentials are determined from data on squared charge radii. Highlighting the case of mesons here for simplicity of discussion, the Wigner distributions $W(\mathbf{x},\mathbf{p})$ of a meson is convoluted with wave packets for the quark and antiquark, assumed to be of generic Gaussian shape. The resulting phase space overlap probability is
\begin{multline}
    P_\text{ps}(\mathbf{r},\mathbf{q}) \sim \int \prod_{i=q,\bar q} d^3 x_i d^3 p_i  W_M(\mathbf{x},\mathbf{p}) \\ \times W_q(\mathbf{x}_1,\mathbf{p}_1) W_{\bar q}(\mathbf{x}_2, \mathbf{p}_2)
\end{multline}
where $\mathbf{x}=\mathbf{x}_1-\mathbf{x}_2$ and $\mathbf{p}=(m_2 \mathbf{p}_1-m_1\mathbf{p}_2)/(m_1+m_2)$ are the relative position and weighted momentum of the two quarks. $m_1$ and $m_2$ are the constituent masses of the quark and antiquark, respectively. The probabilities $P_\text{ps}(\mathbf{r},\mathbf{q})$ depend on the relative phase space coordinates $\mathbf{r}$ and $\mathbf{q}$ of the centroids of the coalescing quark and antiquark wave packets. The latter are, in an abuse of notation, often referred to as \emph{the} positions and \emph{the} momenta of the quark and antiquark, respectively. Note that unlike the Wigner distributions themselves the $P_\text{ps}(\mathbf{r},\mathbf{q})$ are positive definite probabilities fully consistent with quantum mechanics. The caveat is that the correct shape of the wave packets is not known as it is not provided by the shower Monte Carlos. 
The phase space probability is multiplied by the corresponding 
probablities $P_\text{spin}$ and $P_\text{color}$ for overlap of the spin and color states, respectively. The spin of quarks is always treated statistically, i.e. $P_\text{spin}(S=0)=1/4$, $P_\text{spin}(S=1)=3/4$. Information from color tags is used if possible, with noteable cases for mesons being $P_\text{color}=1$ for the coalescence of a quark and antiquark with known equal color/anticolor tags, $P_\text{color}=0$ if  the two color tags are known to not be in a color singlet configuration, and $P_\text{color}=1/9$ for mesons from quarks and antiquarks with random (or zero) color tags.
A similar formalism is in place for baryons \cite{Han:2016uhh}.

The Hybrid Hadronization algorithm considers all possible pairs of quarks and antiquarks and all possible triplets of (anti)quarks to form candidate hadrons, and computes the total probability $P_\text{tot}=P_\text{ps}P_\text{spin}P_\text{color}$ in the center of mass frame of the candidate hadron.
A throw of dice compared to $P_\text{tot}$ decides on the success of the candidate actually forming a hadron. Once a hadron is formed the partons used are stricken from the active parton list. The algorithm continues until all shower partons are either used up, or had all their possible candidate hadrons rejected.

Naturally, partons at a large distance in phase space from other partons, are less likely to find a partner for recombination. This can be observed, e.g., in the high-$z$ core of jets. On the other hand, partons with small momenta, 
in particular those close close in momentum and position to the background bath of partons, are quite likely to directly form hadrons through recombination. The latter can be observed readily in panel (d) of Fig.\ \ref{fig:hadronchannels}.

\subsection{Remnant String Handling and Fragmentation}
\label{subsec:string}

After the recombination step, the partons which have not yet hadronized are gathered in a list of remnant partons.
Since only color singlet configurations have been removed in recombination, the remnant parton system remains in the same color state as before. There are two possibilities. If a complete color singlet event had been input, e.g.\ a complete list of partons with color tags from an $e^+$+$e^-$ collision, the remnant parton system will usually form a string system acceptable to PYTHIA 8 right away.
Otherwise, e.g.\ if cuts to the parton event have been applied, or partons with color tag zero are present, the remnant parton list typically only forms partially reasonable strings.
Strings are then repaired with minimal assumptions  such that color singlets are formed. For this one uses typically partons from the background medium close by in phase space, if available, or beam partons, if there is no medium.
In the current study a single high momentum quark initiates a jet shower and therefore  it is necessary, and sufficient, to add one antiquark to the system at this stage. In this setup we simply assume this parton to have vanishing momentum and the missing color tag.

Note that for each baryon (antibaryon) created by recombination an antijunction (junction) is created in the system of remnant partons. This is the mechanism to conserve net baryon number. Once an acceptable color singlet string system has been achieved,  all partons with their color tag information, and all the information about junctions is transcribed into a PYTHIA 8 event. Hadrons from recombination can be added if one would like to use PYTHIA to decay excited hadrons, as is done here. Hybrid Hadronization then hands the remnant system plus hadrons over to PYTHIA 8 for string fragmentation and hadron decays.

\section{In-Medium Hadronization with the JETSCAPE Framework} 
\label{sec:js_setup}

\begin{table}[tb]
\begin{tabular}{|l|c|}
\hline
\textbf{Setting}     & \textbf{Value} \\ \hline
$t_{\max}$           & 20 fm          \\ \hline
$\alpha_s$ (fixed)   & 0.2           \\ \hline
$Q_0$                & 1.0 GeV        \\ \hline
$T_{\text{brick}}$   & 300 MeV        \\ \hline
$T_c$                & 160 MeV        \\ \hline
N     & 2            \\ \hline
\end{tabular}
\caption{JETSCAPE 3.0 settings used in the study of Hybrid Hadronization: total simulation time, strong coupling constant, lower virtuality cutoff for MATTER, brick temperature, hadronization temperture, allowed hadron excitation level.}
\label{tab:settings}
\end{table}

\begin{figure*}[tb]
    \includegraphics[width=\columnwidth]{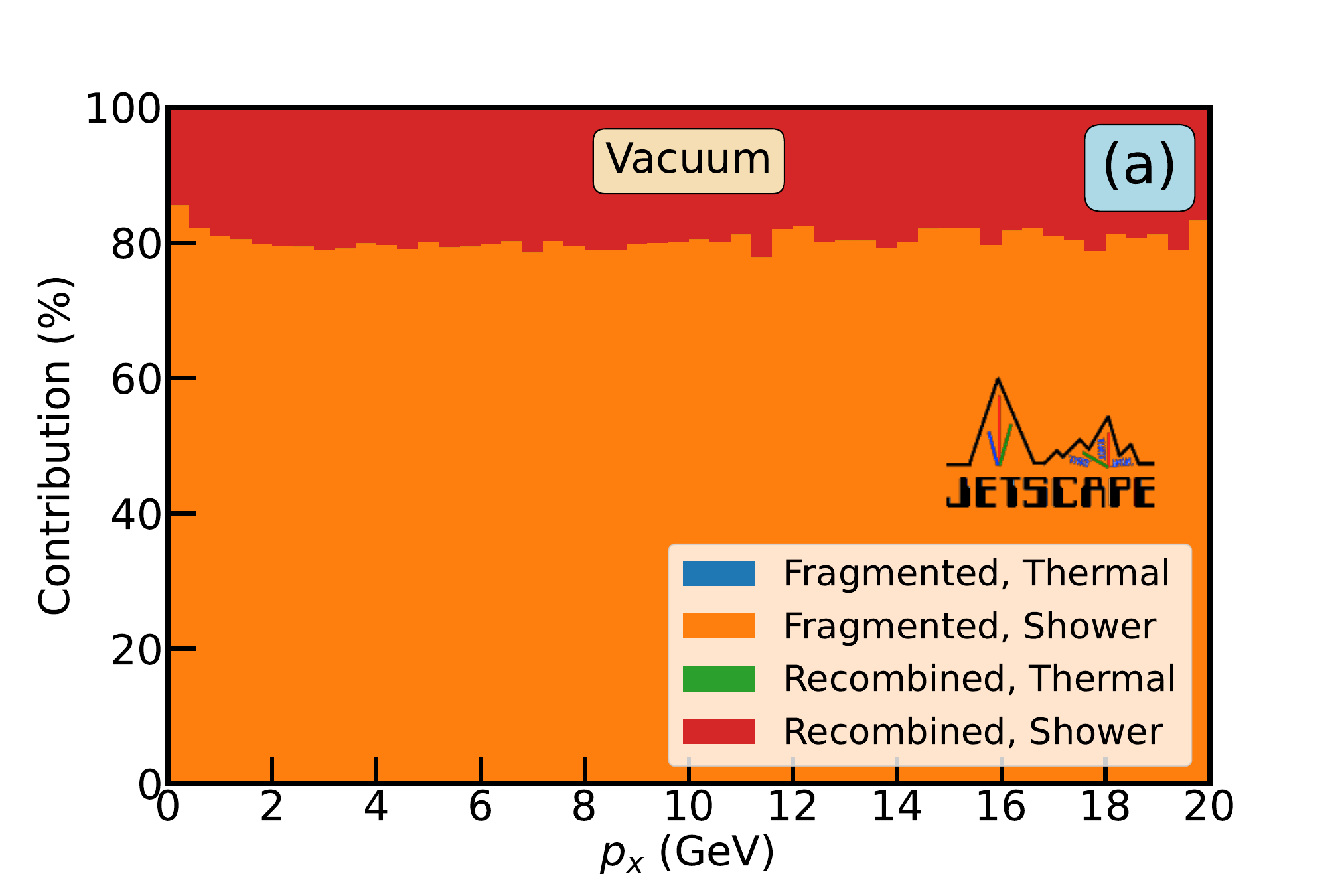}
    \includegraphics[width=\columnwidth]{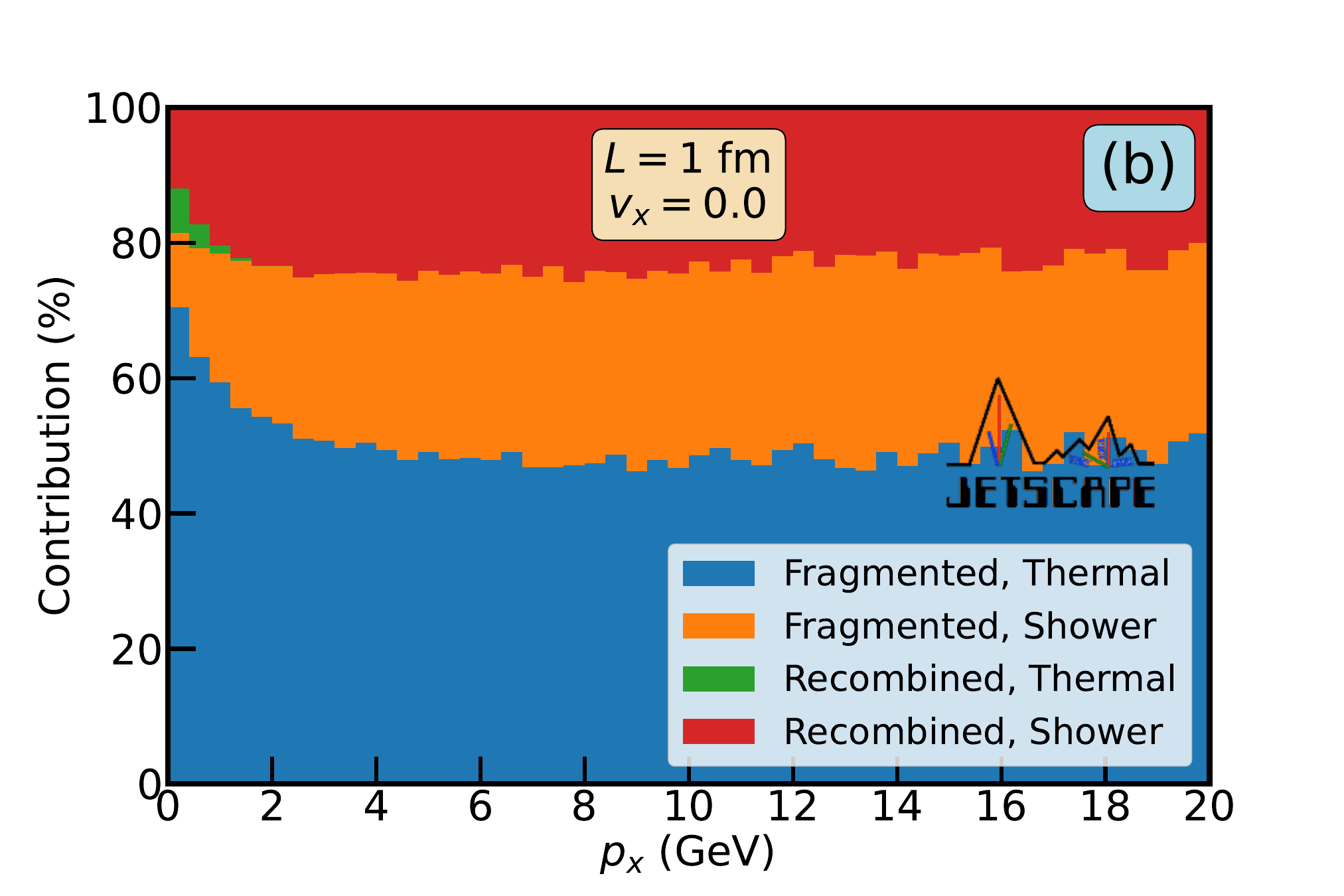}
    \includegraphics[width=\columnwidth]{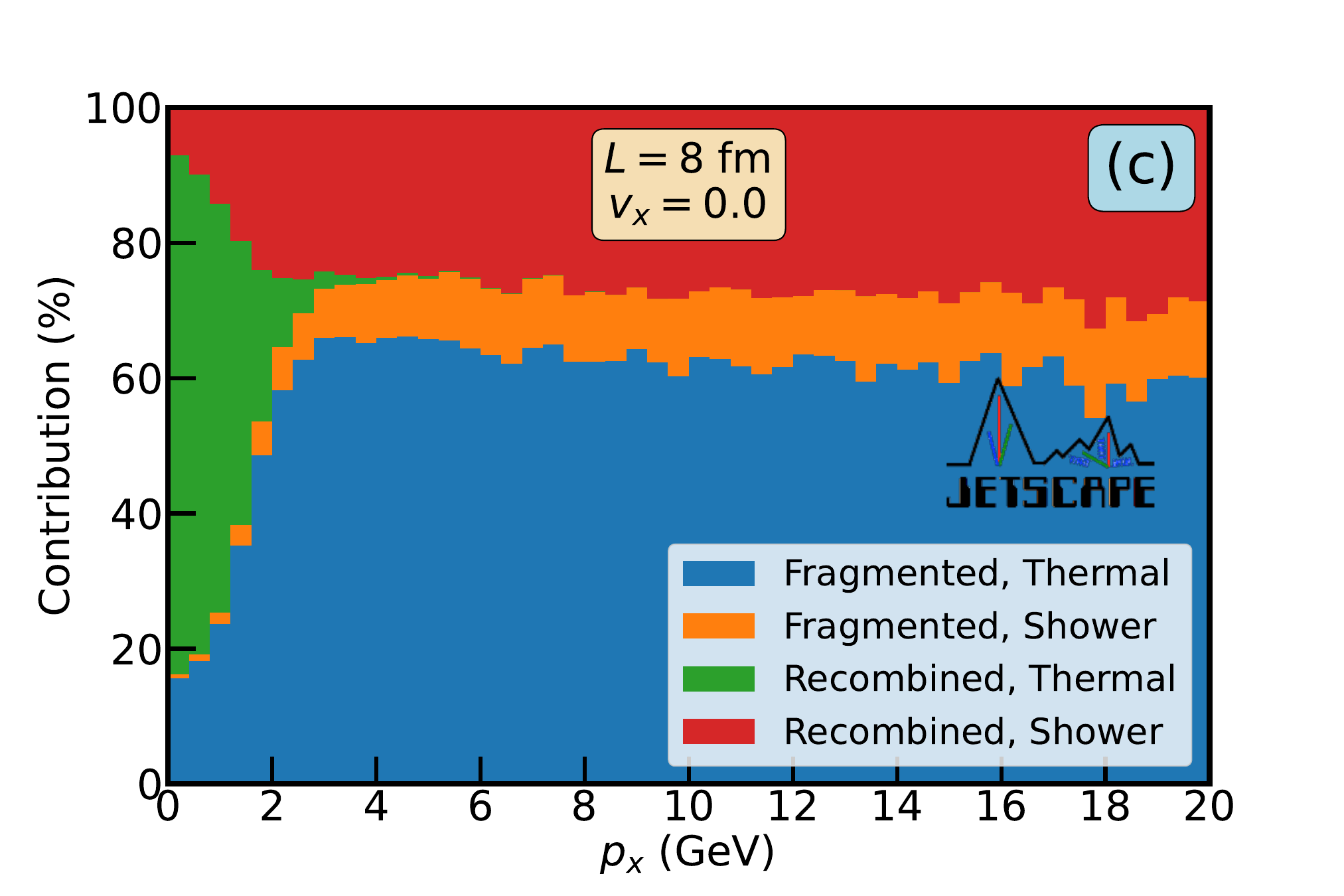}
    \includegraphics[width=\columnwidth]{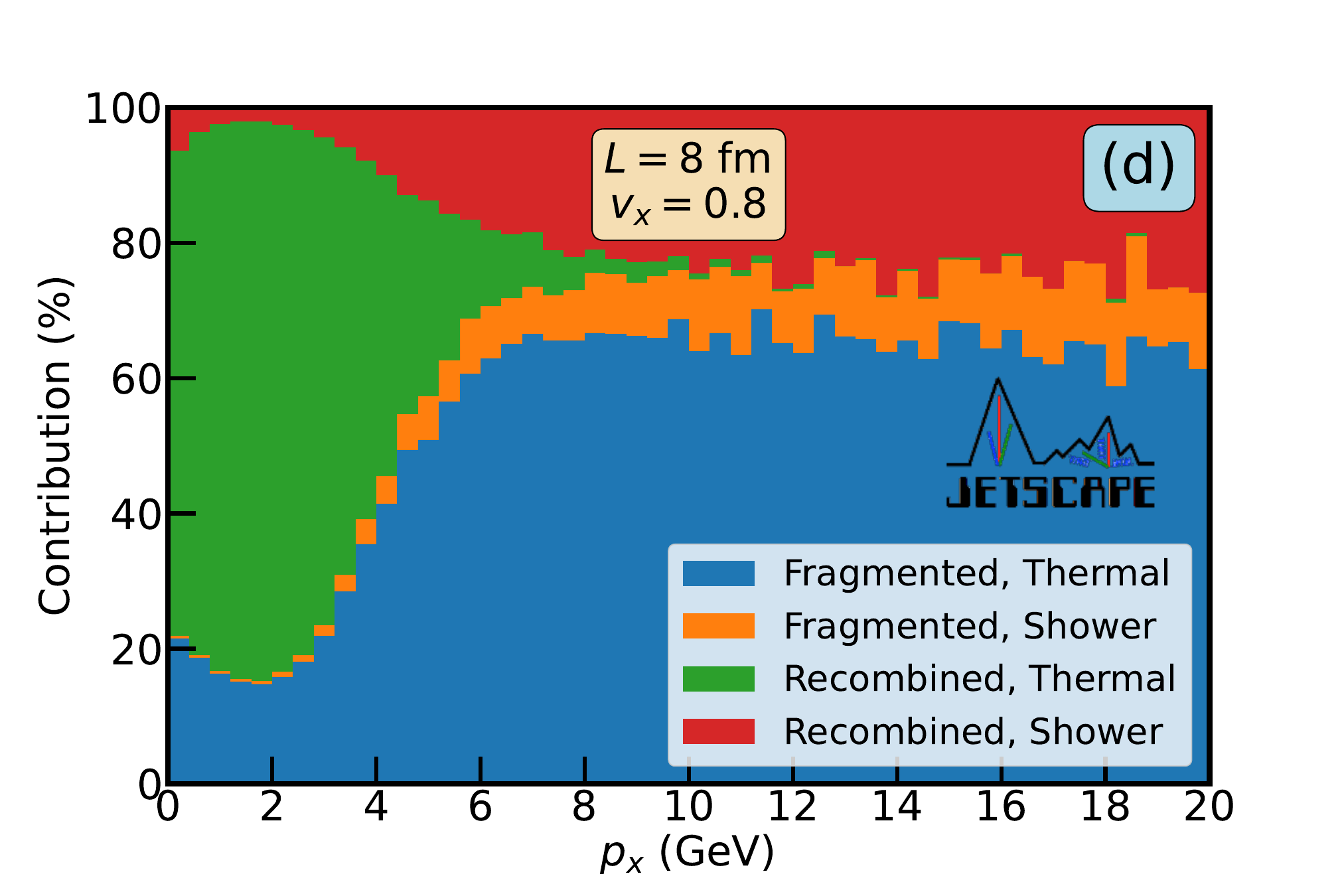}
  \caption{Contributions of different channels to total hadron production numbers from $E=100$ GeV light quark showers as a function of hadron momentum $p_x$ (along the jet): (a) vacuum jets, (b) jets in a small medium of size $L=1$ fm without flow, (c) jets in a large medium ($L=8$ fm) without flow, (d) jets in a large medium ($L=8$ fm) with longitudinal flow $v_x=0.8 c$.}
  \label{fig:hadronchannels}
\end{figure*}

We conduct a systematic study of Hybrid Hadronization within the JETSCAPE framework. This involves studying the effects of medium size and collective flow on observables related to hadron production, hadron spectra, hadro-chemistry and transverse flow.

As a proxy for $A$+$A$ collisions, we use a temporal QGP brick medium whose temporal extent is given by the parameter $L$. The temperature is fixed at $T=300$ MeV (Tab. \ref{tab:settings}) until a time $t=L/c$ is reached, at which the temperature drops to $T_c=160$ MeV for hadronization. Recombination and fragmentation are then carried out until all partons are hadronized. The brick is sufficiently large in all spatial directions to cover the jet. Our simulation is thus akin to a space-like hypersurface with a fixed hadronization time.

We use version 3.0 of JETSCAPE, with settings summarized in Tab. \ref{tab:settings}. The framework is set up as follows. A parton gun (PGun) fires a parton with a given energy and flavor in the $x$-direction, which mimics a parton coming out of a hard process with some virtuality.
MATTER is used to propagate and split the initial parton and the subsequent created partons, generating a jet shower. The initial virtuality of the parton from the PGun is sampled by MATTER with an upper cutoff of half the parton energy. Once the virtuality of a parton falls below a threshold $Q_0$, set here to 1 GeV,  it will no longer be considered by MATTER. 
If this happens beyond a time $L/c$, i.e. outside of the medium, such a parton is immediately ready to hadronize. If a parton's virtuality falls below $Q_0$ before the time $L/c$, i.e. inside the QGP medium, the framework will hand it over to LBT to propagate it further through the medium until time $L/c$.
Eventually, this procedure will lead to an ensemble of real or low-virtuality partons that are ready to hadronize. These partons should either be on the $T=T_c$ hypersurface ( $t=L/c$) or in the vacuum ($t>L/c$).  At this point the framework hands the list of partons over to Hybrid Hadronization.

Hybrid Hadronization runs as described in the previous section. Two levels of excited states are considered beyond the ground states ($N=2$, see Ref.\ \cite{Han:2016uhh}). The cutoff for the decay of resonances is set to 1 cm$/c$  in PYTHIA. Hybrid Hadronization tags the origin of hadrons as either recombination or fragmentation. It also tracks if a recombined hadron, or a string contains any thermal partons.

In some of the results below, the Liquefier component of JETSCAPE is used to remove very soft partons from the jet shower \cite{JETSCAPE:2020uew} as a very simple background subtraction procedure. Soft partons are defined here as having an energy $<3.2 \,T$ where $T$ is the local temperature in the ambient fluid rest frame. In a setup in which the QGP medium evolves dynamically with the jet shower this procedure could in principle be used to properly account for energy and momentum flowing between the medium and the jet shower. This is not done in the simple case of a static medium here. Rather, the scale of $3.2 \,T$ is chosen to roughly recover the original jet energy. The Liquefier is only used here where explicitly stated.

\section{Results} 
\label{sec:results}
We now proceed to show the results from our setup. First, we check whether recombination and fragmentation exhibit their expected behavior in terms of their relative importance as a function of both medium size and medium flow. It is expected that hadrons at low and intermediate $p_T$ are more likely to come from recombination and we thus focus our analysis on hadrons with momentum (along the jet direction) of 10 GeV/$c$ and less in most cases. Hadrons from thermal-shower recombination in particular are expected to be limited in their range in $p_T$ due to the softness of the thermal parton spectrum.

We then proceed to investigate the effects of in-medium hadronization on observables at low and intermediate $p_T$, in particular the ratios of in-medium and vacuum fragmentation functions, and the ratios of baryons over mesons. The expectation is that recombination in a medium is able to produce baryons without penalty compared to mesons \cite{Fries:2008hs}. This contrasts with the vacuum case where, starting from a $q\bar q$ string, baryon formation requires an additional quark-antiquark pair to be created compared to the meson case. The baryon/meson suppression can, e.g.,  be observed in $e^+$+$e^-$ collisions at the $Z$-pole, where the ALEPH collaboration reports $p/\pi\sim 0.1$ for $p_T$ around 2 GeV$/c$  \cite{ALEPH:1994cbg}. On the other hand, in nuclear collisions the proton/pion ratio reaches $\sim 1$ around $p_T\sim 3-4$ GeV$/c$ in central collisions \cite{PHENIX:2013kod,ALICE:2019hno}. Attempts have been made to attribute this enhancement to purely thermal production of hadrons at 3-4 GeV/$c$ \cite{Hirano:2003pw}. However, the constituent quark scaling of elliptic flow observables at the same hadron $p_T$ \cite{STAR:2008ftz,ALICE:2014wao} is challenging to explain in the latter picture, while it naturally by quark recombination \cite{Fries:2008hs}. We emphasize once more that soft hadrons, i.e. any hadrons not associated with the jet, are not added to the final hadron spectra. Therefore, our results cannot be directly compared to single inclusive measurements of hadrons in $p$+$p$ or $A$+$A$ collisions, in particular in the range below 2-3 GeV/$c$.

\begin{figure*}[tb]
	\includegraphics[width=\linewidth]{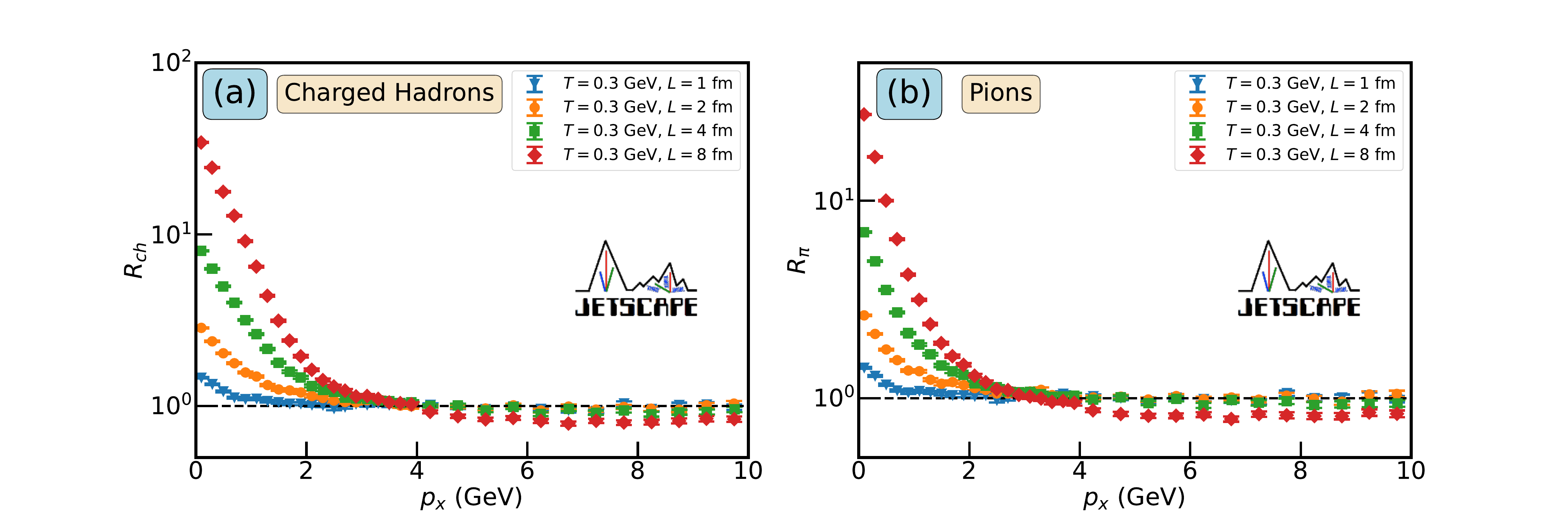}
	\includegraphics[width=\linewidth]{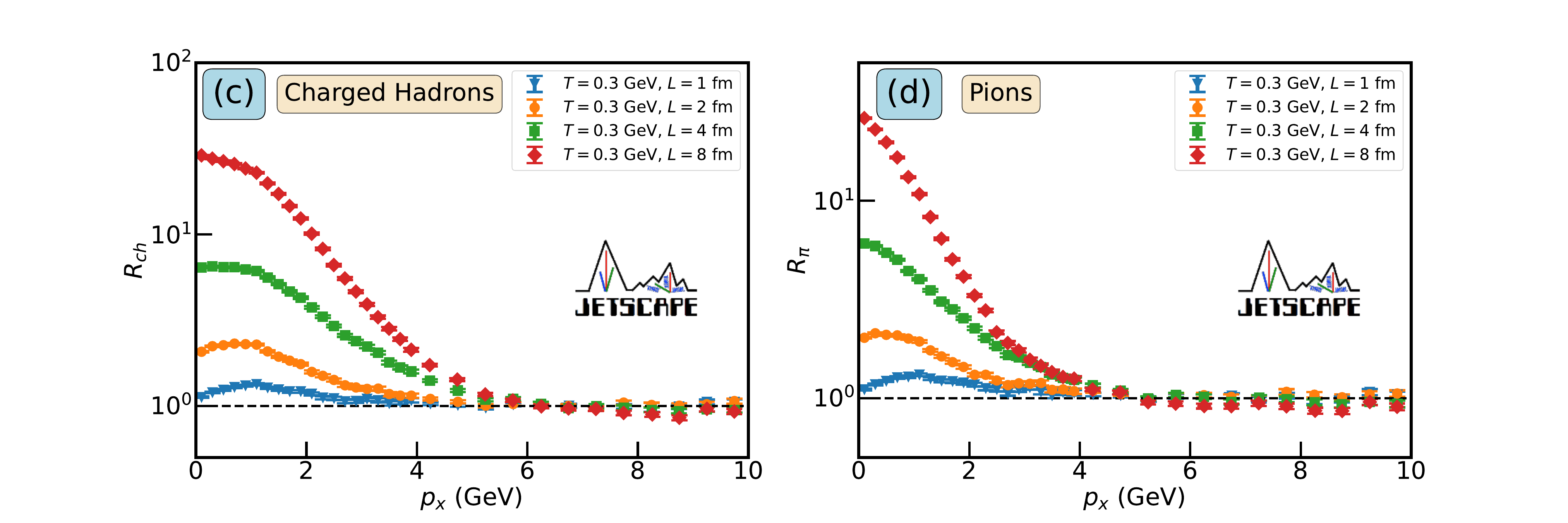}
  \caption{Ratios of fragmentation functions $[dN/dp_x]_\text{medium}/[dN/dp_x]_\text{vac}$ for charged hadrons (left panels) and pions (right panels) from $E=100$ GeV light quark showers, in media of various lengths $L$ from 1 to 8 fm. Top panels: no ambient medium flow. Bottom panels: medium with flow longitudinal to the jet, $v_x=0.8c$. }
  \label{fig:fragfunc}
\end{figure*}

Finally, we explore how medium flow transverse to the jet direction has effects distinct from those of longitudinal flow. In particular, we study the different effects on soft hadrons and jet shape. Generally, we expect the interaction of jet partons with a thermal medium to lead to diffusion in the transverse direction. This effect should be visible in the jet shape. A preferred medium flow in the transverse direction should introduce an additional bias of hadron momenta in the same direction. On the other hand, the main effect of longitudinal flow should be an increased amount of recombination as more partons at intermadiate $p_T$ of a few GeV$/c$ have a chance to find a velocity-matched thermal parton to coalesce with.

\begin{figure*}[tb]
	\includegraphics[width=0.9\linewidth]{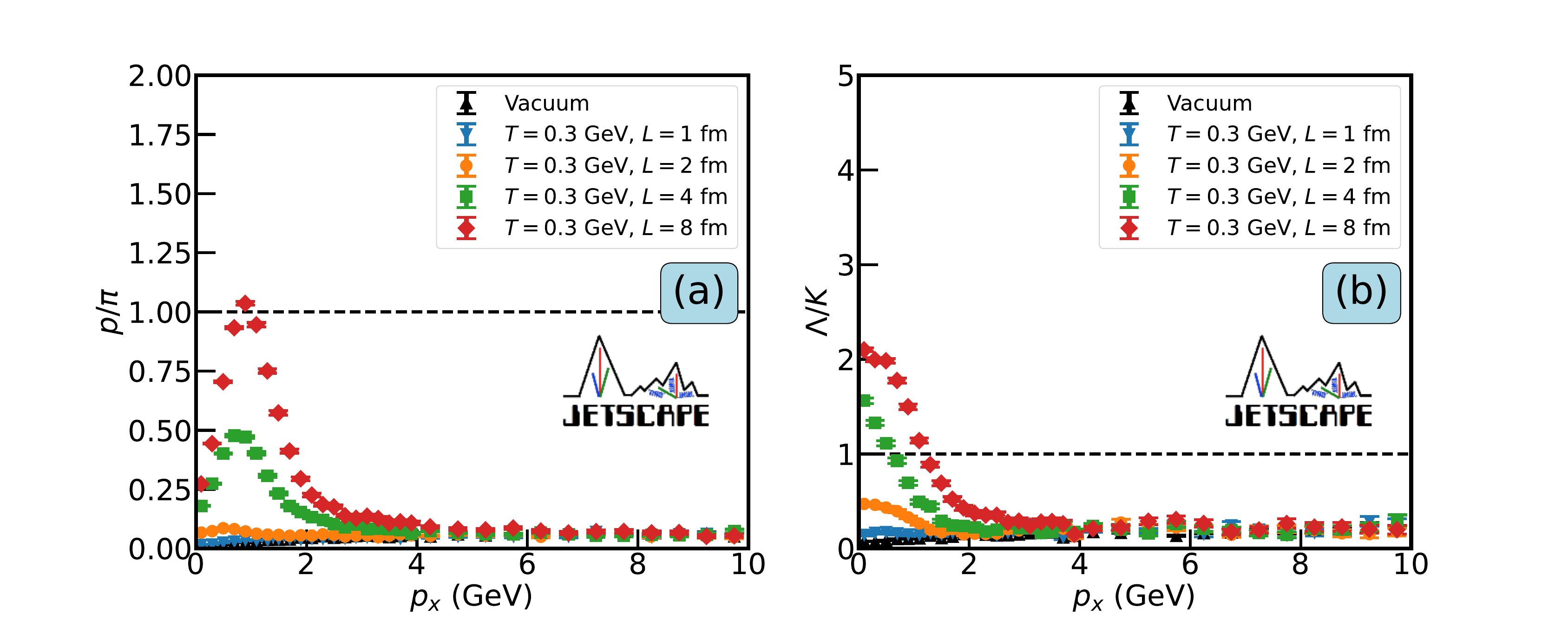}
	\includegraphics[width=0.9\linewidth]{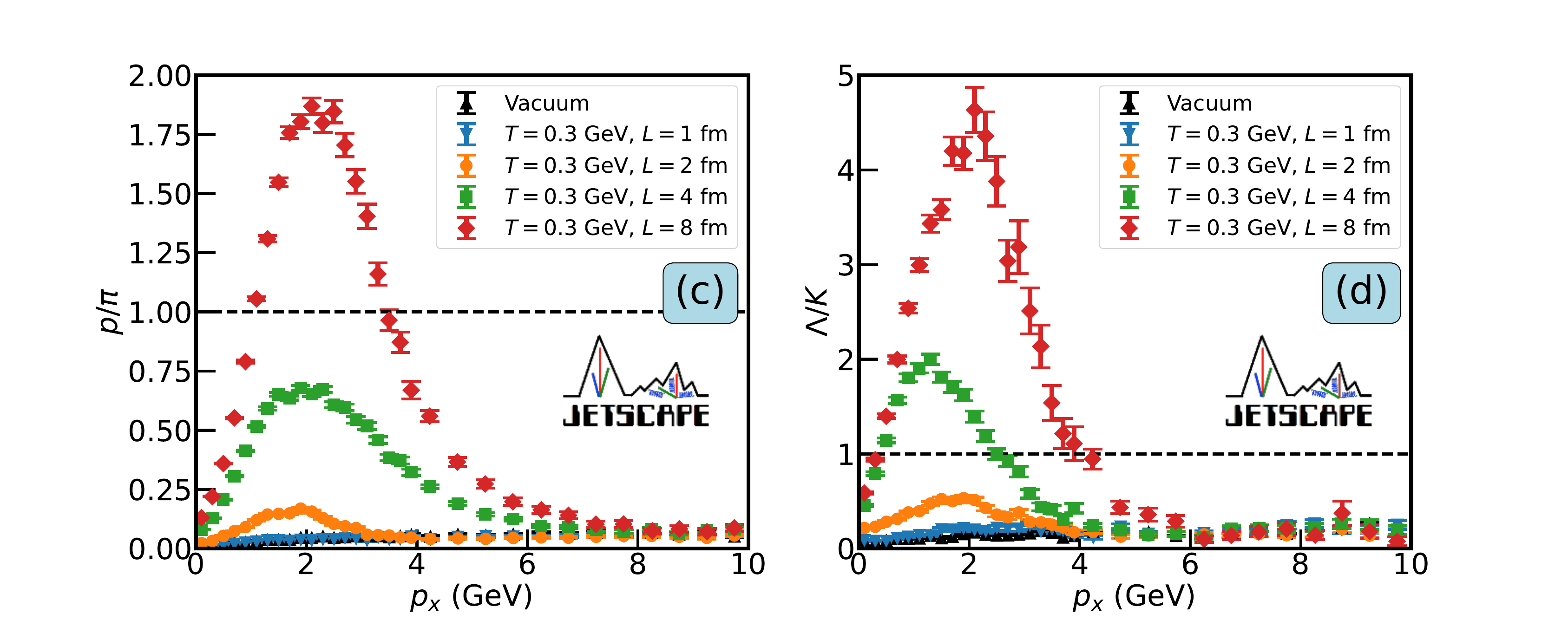}

  \caption{Proton/pion (left panels) and $\Lambda$/kaon ratios as functions of $p_x$ for $E=100$ GeV light quark showers, in vacuum and for medium sizes $L$ from 1 to 8 fm. Top panels: no ambient medium flow. Bottom panels: medium with flow longitudinal to the jet, $v_x=0.8c$.}
  \label{fig:hadrochem}
\end{figure*}

\subsection{Hadron Production Channels}
\label{subsec:contrib}

For jet showers initiated by a light quark with energy $E=100$ GeV Fig.\ \ref{fig:hadronchannels} shows the contributions of four different hadron production channels to the total number of hadrons for a given longitudinal (to the jet) momentum $p_x$. 
The top left panel shows the baseline jet without a medium, i.e. only MATTER with vacuum splittings is active in the partonic phase, and Hybrid Hadronization does not include any thermal partons. We note that shower-shower recombination effects (red shaded) are active even in the vacuum but contribute less than 20\% of the observed hadron yield. Interestingly, we find the contribution from recombination is rather stable up to 20 GeV/$c$. Note that recombination processes in a highly boosted object like a jet shower always push particles to higher momenta.

Next, we want to confirm that shower-thermal recombination switches on smoothly when going from small to large systems. This is a basic requirement that Hybrid Hadronization needs to clear. The upper right and lower left panels of Fig.\ \ref{fig:hadronchannels} show hadrons when the same jets are embedded in media of sizes of 1 fm and 8 fm respectively. For the smaller medium we notice a small number of hadrons from shower-thermal recombination at very low momenta. The importance of the shower-thermal recombination channel grows significantly in the larger medium where it is the most important channel below $2$ GeV/$c$. We have run simulations for other medium sizes (not shown here) which confirm a smooth turn on.

We also break up the contributions from remnant string fragmentation into a channel with only shower partons in the string producing a hadron, and a channel for hadrons with at least one thermal parton in a string. The latter channel jumps to a large value already for a small medium. This is due to the fact that our single jet-system is not a color singlet and always picks at least one thermal parton to make all strings color singlets. If one takes a very peripheral $A$+$A$ collision as an example, the color singlet condition might still be satisfied otherwise, e.g. by connecting the string to a beam remnant. We can therefore not draw strong conclusions from the distinction of the two recombination channels. For larger systems this is likely different as most of the jet will spend a significant amount of time inside the hot parton medium, erasing the existing color flow. One important conclusion from the result for $L=8$ fm is thus the persistence of a pure shower fragmentation channel. This reflects the underlying space-time picture in which the leading part of a 100 GeV jet easily extends beyond the medium due to time dilation. A significant amount of radiation can still happen in this jet "core", creating partons and a jet-like color flow outside the medium.

Panel (d) of Fig. \ref{fig:hadronchannels} shows the effect of medium flow in the direction of the jet. This simulates radial flow in an $A$+$A$ collision parallel to the trajectory of a jet. We only show the case of $L=8$ fm and $v_x=0.8 c$ here. Note that in our study collective flow is only given to the thermal partons for hadronization. It is not active during the MATTER and LBT phases, to clearly demonstrate flow effects from in-medium hadronization. Our calculations thus estimates a lower bound for full flow effects in which flow already builds up during the parton phase.

We find that longitudinal flow greatly enhances the prevalence of shower-thermal recombination, making it dominant up to $\sim 4$ GeV$/c$ and still contributing to the total hadron yield out to about $\sim 8$ GeV$/c$. Once again this effect switches on smoothly as $v_x$ is varied from 0 to 0.8$c$.

\begin{figure*}[tb]
	\includegraphics[width=\columnwidth]{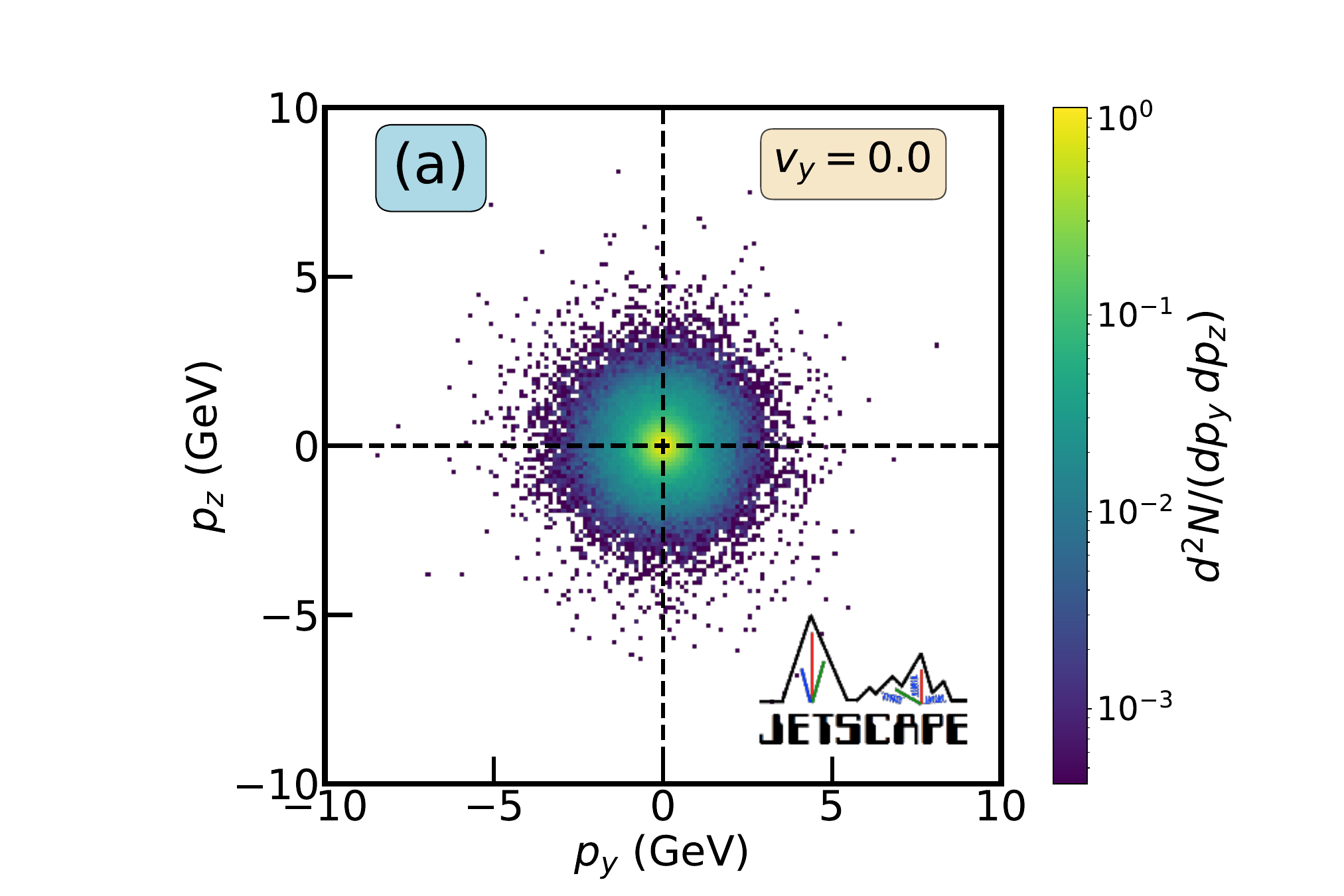}
    \includegraphics[width=\columnwidth]{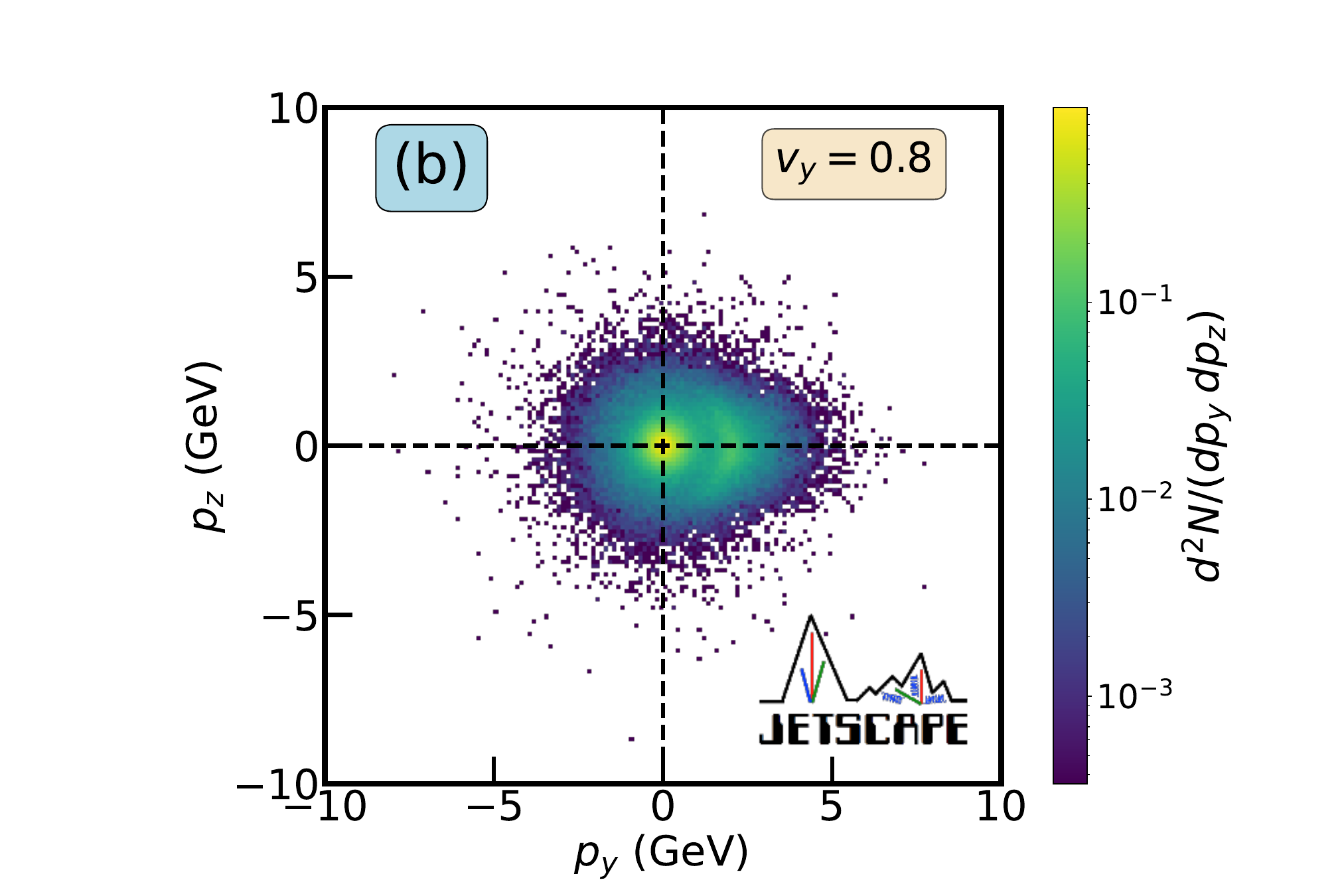}
  \caption{Transverse distribution, $1/N^{\text{jet}}\sum_{\text{jets}}d^2N/(dp_y\:dp_z)$, of intermediate (2-10 GeV$/c$) hadrons from $E=100$ GeV light quark showers in a $L=4$ fm medium: (a) medium without flow, (b) medium flow transverse to the jet, $v_y=0.8$.}
  \label{fig:transflow_soft}
\end{figure*}
\begin{figure*}[tb]
	\includegraphics[width=\columnwidth]{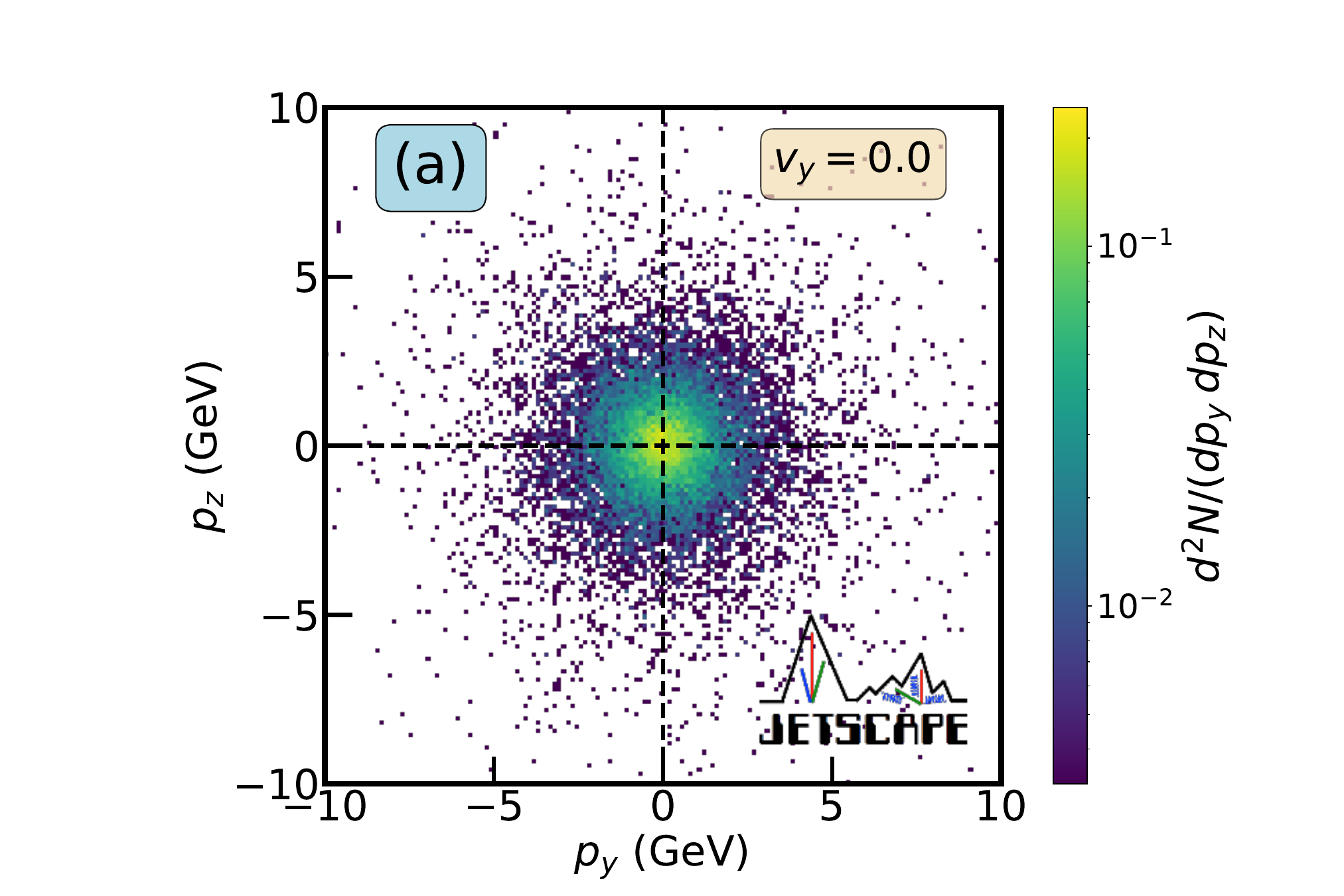}
    \includegraphics[width=\columnwidth]{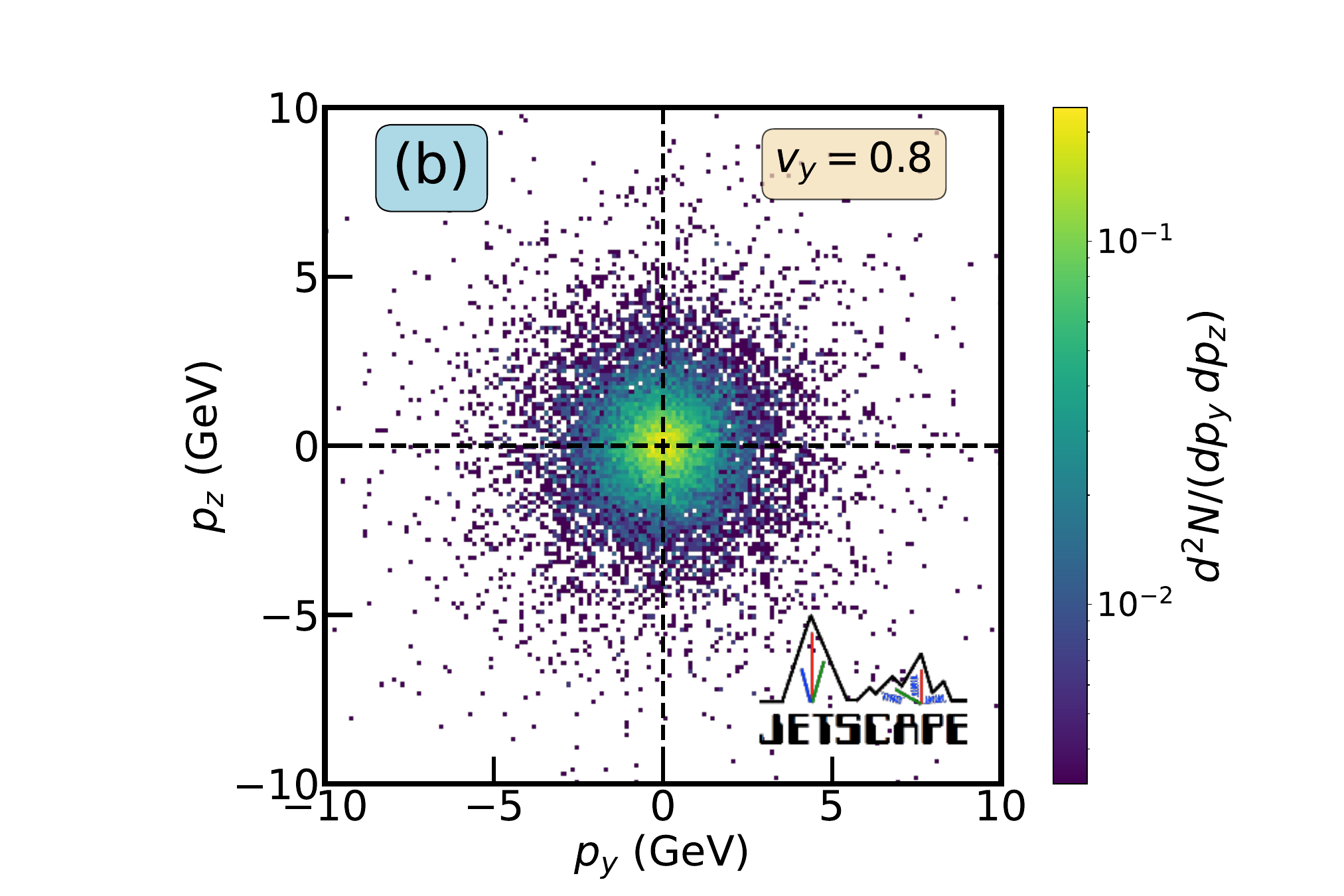}
  \caption{Same as Fig. \ref{fig:transflow_soft} for the leading jet hadrons.}
  \label{fig:transflow_hard}
\end{figure*}

\subsection{Hadron Spectra}
\label{subsec:reco}

Parton spectra before hadronization are modified for jets in a medium compared to vacuum jets. This effect is computed in our setup by MATTER and LBT and it is at the center of many competing calculations in the literature \cite{Wang:2001ifa,Gyulassy:2003mc,Salgado:2003gb,Qin:2007rn,JET:2013cls,Casalderrey-Solana:2014bpa} Quenching of the parton shower results in a reduction of partons at large momenta and an accumulation of additional partons at lower momenta. The distinction between the jet and the medium can become unclear as partons from the thermal medium can be scattered to become part of the jet shower, while shower partons that loose sufficient energy thermalize into the medium.
In-medium hadronization naturally continues this process as medium partons become part of hadrons in the jet shower.  We therefore expect both shower-thermal recombination and strings connecting to thermal partons to influence hadron spectra at low and intermediate momenta. 

We investigate the size of the combined effects of parton quenching and in-medium hadronization by reporting the hadron fragmentation function ratio $R_{AA}$, defined here as
\begin{equation}
    R_{AA} = 
    \dfrac{\left[dN/dp_x\right]_{\text{medium}}}{\left[dN/dp_x\right]_{\text{vacuum}}}
\end{equation}
for charged hadrons and pions in media of varying size and longitudinal flow. Fig.\ \ref{fig:fragfunc} shows results for $E=100$ GeV light quark showers. The size $L$ of the medium is  varied between 1 and 8 fm, and results are presented with or without longitudinal flow.  

In panels (a) and (b) we observe an increase of the amount of soft hadrons (below 2 GeV$/c$)  compared to the vacuum case even for small media, as expected. The excess of soft partons is significantly larger for larger media, seemingly growing faster than linear with $L$. We also see jet quenching at values of $p_x$ higher than 4 GeV$/c$, which grows with medium size. Recall that we only count hadrons associated with the jet shower, i.e.\ we exclude hadrons that only contain partons from the background medium. The in-medium spectra therefore include excess energy and momentum imparted by the medium which would have to be subtracted from the medium in a full simulation.

Collective flow of the medium longitudinal to the jet shifts the enhancement of hadrons to larger momenta. This is shown in the bottom panels of Fig.\ \ref{fig:fragfunc} for $v_x=0.8c$. 
Significant enhancement of fragmentation functions is still seen around 4 GeV$/c$ for charged hadrons. The shoulder developing at low $p_x$ for charged hadrons but not for pions is a signal for collective motion in which similar velocities for different masses lead to different momenta.

\begin{figure*}[tb]
	\includegraphics[width=\columnwidth]{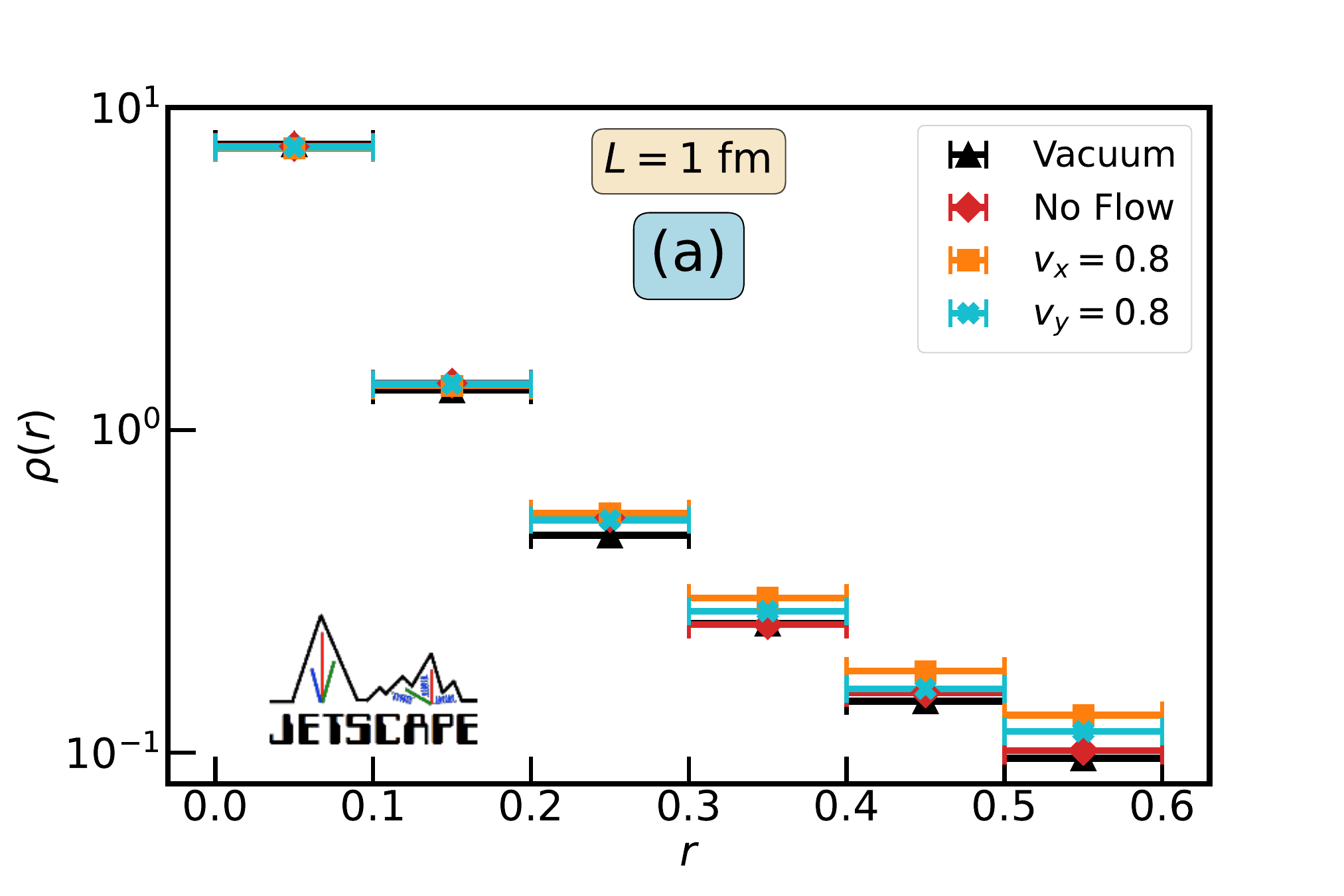}
    \includegraphics[width=\columnwidth]{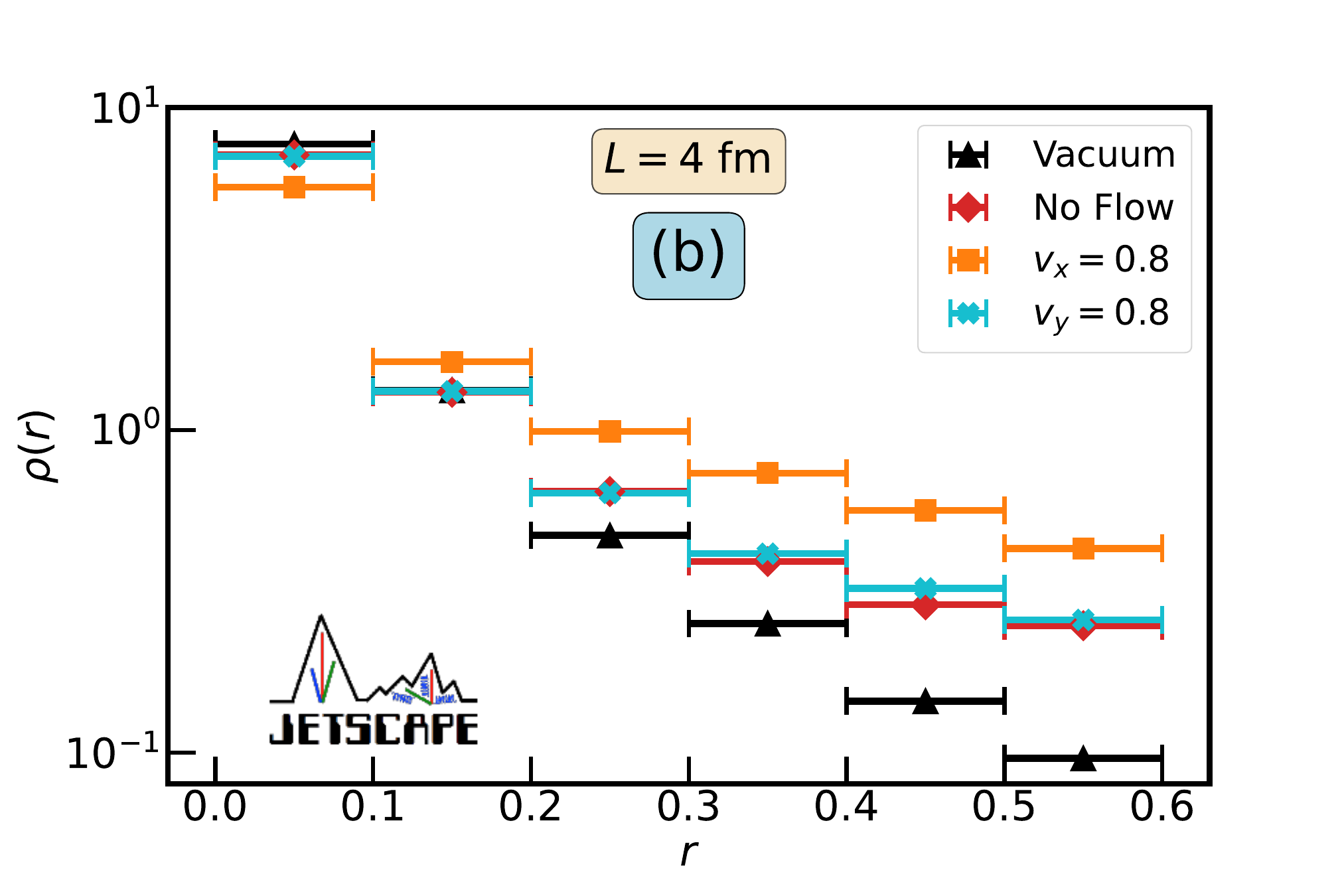}
  \caption{Transverse jet shape $\rho(r)$ for $R=0.6$ anti-$k_T$ jets in vacuum and in medium, without flow and with either longitudinal or transverse flow: (a) $L=1$ fm, (b) $L=4$ fm.}
  \label{fig:transjetshape}
\end{figure*}

\subsection{Hadro-Chemistry}
\label{subsec:chemistry}

Baryon/meson ratios are considered a hallmark signature of recombination since baryon production is not penalized by mass or number of valence quarks. 
The top panels of Fig. \ref{fig:hadrochem} show $(p+\bar p)/(\pi^++\pi^-)$ and ($\Lambda+\bar{\Lambda})/(K^++K^-)$ ratios for $E=100$ GeV light quark showers for vacuum jets and various medium sizes from 1 to 8 fm with and without flow longitudinal to the jet. Our results show a monotonous increase of both of these ratios with vacuum results close to the expected values and the $p/\pi$ ratio being close to unity for the largest medium. For a background medium without flow the enhancement peaks for soft momenta $p_x \lesssim 2$ GeV$/c$. Clearly, the growing share of shower-thermal recombination seen in Fig.\ \ref{fig:hadronchannels} is responsible for this behavior.

The bottom panels of Fig.\ \ref{fig:hadrochem} show baryon/meson ratios with collective medium flow along the direction of the jet with $v_x=0.8c$. As seen for fragmentation functions, recombining hadrons pick up additional momentum from the medium, which shifts the peaks in the ratios towards intermediate momenta. The height of the peaks also increases as baryons receive a further flow push on spectra that are steeply falling. Again, the existence of a flow signal in jet hadrons is a direct reflection of flow present in the medium.
Hybrid hadronization successfully models the hadrochemistry effects expected from in-medium hadronization.

\subsection{Transverse Flow Effects and Jet Shapes}
\label{subsec:flow}

For the remainder of this section we look at effects that have not been discussed much in the literature in terms of in-medium hadronization. First, we investigate how collective flow in the ambient medium \emph{transverse} to the jet propagation axis modifies hadron distributions transverse to the jet. Such flow can arise in various scenarios.

Figs.\ \ref{fig:transflow_soft} and \ref{fig:transflow_hard} show distributions
\begin{equation*}
    \dfrac{1}{N^{\text{jet}}}\sum_{\text{jets}}\dfrac{d^2N}{dp_y\:dp_z}
\end{equation*}
in the plane transverse to the jet for intermediate hadrons, defined to have $p_x$ between 2 and 10 GeV$/c$, and for the hardest parton in a jet, respectively. The medium is chosen to have size $L=4$ fm with no longitudinal flow. The transverse flow is zero in panels (a) and is set to $v_y=0.8c$ (in the $y$-direction) in panels (b).

Without transverse flow both leading and intermediate  hadrons are centered around the origin in the transverse momentum plane, as expected, with no preferred direction. The presence of transverse flow during hadronization induces a deformation of intermediate momentum hadrons in the direction of the flow vector, in this case the $y$-axis. On the other hand, no significant  deformation is visible for leading hadrons in the presence of flow. 
This is yet another manifestation of the fact that soft and intermediate momentum partons in the jet are likely to find matching thermal partons in phase space and can thus receive transverse momentum kicks from the medium. The emerging hadrons carry additional momentum in the preferred direction. At the same time, leading hadrons are unlikely to arise from thermal-shower recombination and do not show signs of flow. Thus Hybrid Hadronization 
leaves jet cores largely unaffected by medium effects. 
Note that the Liquefier has been used for this transverse flow study to reduce background contamination.

One standard observable for studying the transverse jet structure is the transverse jet shape $\rho(r)$ defined as \cite{pppaper}
\begin{equation}
    \rho(r) = \dfrac{1}{N^{\text{jet}}}\sum_{\text{jets}}\left[\dfrac{1}{\delta r} \dfrac{\sum_{i\in(r\pm\delta r/2)}p_T^i}{\sum_{i\in(0,R)}p_T^i}\right]
    \label{eq:rho}
\end{equation}
for a jet clustered with radius $R$ and where $r = \sqrt{(\Delta\eta)^2 + (\Delta\phi)^2}$ is the distance in the rapidity-azimuth plane. $\rho(r)$ represents the fraction of jet-$p_T$ in the annulus of width $\delta r$, centered about the jet axis. 
In the following, we use the largest energy $R=0.6$ jet in the event, clustered using the anti-$k_T$ algorithm in FastJet \cite{Cacciari:2011ma}.

Results are shown in Fig.\ \ref{fig:transjetshape} for a small medium ($L=1$ fm, panel (a)) and a larger medium ($L=4$ fm, panel (b)). The presence of a medium broadens jets beyond $r=0.2$ compared to the  vacuum case by depleting the amount of momentum $p_T$ at smaller radii and it createsa a corresponding increase at larger radii. This effect is expected due to the additional momentum kicks from in-medium hadronization. The broadening is clearly more pronounced in the larger medium.

Curiously, transverse flow present during hadronization does not significanly change the jet shape $\rho(r)$. However, from the previous discussion of this subsection it is clear that the main effect of transverse flow of the background medium is a dipole-like deformation, rather than an azimuthally symmetric one. Therefore, perhaps it should not come as a complete surprise that its effects on jet shape seem to be small.

Interestingly, longitudinal flow (orange) widens the jet more than transverse flow (green) both for an $L=1$ fm medium and the larger $L=4$ fm medium. 
At first glance, one may expect longitudinal flow to have a focusing, rather than broadening, effect. An explanation can be found in Fig.\ \ref{fig:hadronchannels}. Longitudinal flow significantly increases the amount of recombination with thermal partons at low and intermediate momentum. We therefore expect more broadening in the transverse plane of the jet, due to an increase in transverse momentum kicks provided by boosted thermal partons.

\section{Summary and discussion}
\label{sec:discussion}

Hybrid Hadronization needs to pass two major test before it can claim to be a comprehensive  hadronization models. First, it should be able to hadronize jet showers \emph{in vacuum} to a level of accuracy comparable to state-of-the art models. This is made possible by Hybrid Hadronization referring partons preferentially to string fragmentation in such situations. A quantiative test using data from $e^++e^-$ and $p+p$ collisions will be reported on elsewhere. Note that any hadronization model for jet showers can only be tested and tuned in a meaningful way in conjunction with a shower Monte Carlo and a hard process event generator. The second test should confirm that (i) the signature effects expected from hadronization in a medium are present, that (ii) they switch on smoothly with increasing medium size, and that (iii) they are confined to the momentum range in which such modifications are observed in experiment. 

Our work confirms that shower-thermal recombination switches on with the presence of a medium and scales up with increased medium size. Flow of medium partons along the jet axis can bring shower-thermal hadron recombination to the intermediate momentum regime. In larger media, flow parallel to the jet is imprinted quite clearly onto the resulting jet hadrons. Moreover, shower-thermal recombination alters the hadron chemistry in the expected way. Enhanced baryon-to-meson ratios can be observed to grow with medium size, peaking around 1 in large media. If medium flow is taken into account, peak ratios occur at values of $p_T$ similar to those observed in experimental data, although our calculations, lacking soft hadrons, cannot be compared directly to data.

We have also investigated the effect of transverse flow present during hadronization. While intermediate momentum hadrons receive transverse momentum kicks in the flow direction, this is not visible as additional broadening in jet shape. Leading jet hadrons are unaffected by transverse flow during hadronization, as expected.
On the other hand, we observe that longitudinal flow during hadronization had an additional broadening effect on jet shape, likely due to the increase in the overall number of hadrons involving thermal partons.

This study establishes a baseline for in-medium hadronization effects. The next step will be to combine jet shower Monte Carlo simulations with realistic background media computed with fluid dynamics. Such large scale calculations are available but have not yet looked at intermediate momenta and signatures of in-medium hadronization \cite{JETSCAPE:2022jer,JETSCAPE:2023ikg}. These studies will provide the opportunity to test in-medium effects quantitatively, and to explore effects not accessible in a brick study, e.g.\ the scaling of elliptic flow.



\section*{Acknowledgments} 


This work was supported in part by the National Science Foundation (NSF) within the framework of the JETSCAPE collaboration, under grant number \rm{OAC-2004571} (CSSI:X-SCAPE). It was also supported under  \rm{PHY-1516590}, \rm{PHY-1812431} and \rm{PHY-2413003} (R.J.F., M.Ko., C.P., B.K. and A.S.); it was supported in part by the US Department of Energy, Office of Science, Office of Nuclear Physics under grant numbers \rm{DE-AC02-05CH11231} (X.-N.W. and W.Z.), \rm{DE-AC52-07NA27344} (A.A., R.A.S.), \rm{DE-SC0013460} (A.K., A.M., C.Sh., I.S., C.Si and R.D.), \rm{DE-SC0021969} (C.Sh. and W.Z.), \rm{DE-SC0024232} (C.Sh. and H.R.), \rm{DE-SC0012704} (B.S.), \rm{DE-FG02-92ER40713} (J.H.P. and M.Ke.), \rm{DE-FG02-05ER41367} (C.Si, D.S. and S.A.B.), \rm{DE-SC0024660} (R.K.E), \rm{DE-SC0024347} (J.-F.P. and M.S.). This work was also supported in part by the National Science Foundation of China (NSFC) under grant numbers 11935007, 11861131009 and 11890714 (Y.H. and X.-N.W.), by the Natural Sciences and Engineering Research Council of Canada (C.G., S.J., X.W. and G.V.),  by the University of Regina President's Tri-Agency Grant Support Program (G.V.), by the Canada Research Chair program (G.V. and A.K.) reference number CRC-2022-00146, by the Office of the Vice President for Research (OVPR) at Wayne State University (Y.T.), by JSPS KAKENHI Grant No.~22K14041 (Y.T.), and by the S\~{a}o Paulo Research Foundation (FAPESP) under projects 2016/24029-6, 2017/05685-2 and 2018/24720-6 (M.L.). C.Sh., J.-F.P. and R.K.E. acknowledge a DOE Office of Science Early Career Award. I.~S. was funded as part of the European Research Council project ERC-2018-ADG-835105 YoctoLHC, and as a part of the Center of Excellence in Quark Matter of the Academy of Finland (project 346325). Portions of this research were conducted with resources provided by Texas A\&M High Performance Research Computing. 



\bibliography{HybridHad_JS}

\begin{thebibliography}{35}
\expandafter\ifx\csname natexlab\endcsname\relax\def\natexlab#1{#1}\fi
\expandafter\ifx\csname bibnamefont\endcsname\relax
  \def\bibnamefont#1{#1}\fi
\expandafter\ifx\csname bibfnamefont\endcsname\relax
  \def\bibfnamefont#1{#1}\fi
\expandafter\ifx\csname citenamefont\endcsname\relax
  \def\citenamefont#1{#1}\fi
\expandafter\ifx\csname url\endcsname\relax
  \def\url#1{\texttt{#1}}\fi
\expandafter\ifx\csname urlprefix\endcsname\relax\def\urlprefix{URL }\fi
\providecommand{\bibinfo}[2]{#2}
\providecommand{\eprint}[2][]{\url{#2}}

\bibitem[{\citenamefont{Sjostrand et~al.}(2006)\citenamefont{Sjostrand, Mrenna,
  and Skands}}]{Sjostrand:2006za}
\bibinfo{author}{\bibfnamefont{T.}~\bibnamefont{Sjostrand}},
  \bibinfo{author}{\bibfnamefont{S.}~\bibnamefont{Mrenna}}, \bibnamefont{and}
  \bibinfo{author}{\bibfnamefont{P.~Z.} \bibnamefont{Skands}},
  \bibinfo{journal}{JHEP} \textbf{\bibinfo{volume}{05}}, \bibinfo{pages}{026}
  (\bibinfo{year}{2006}), \eprint{hep-ph/0603175}.

\bibitem[{\citenamefont{Sjostrand et~al.}(2015)\citenamefont{Sjostrand, Ask,
  Christiansen, Corke, Desai, Ilten, Mrenna, Prestel, Rasmussen, and
  Skands}}]{Sjostrand:2014zea}
\bibinfo{author}{\bibfnamefont{T.}~\bibnamefont{Sjostrand}},
  \bibinfo{author}{\bibfnamefont{S.}~\bibnamefont{Ask}},
  \bibinfo{author}{\bibfnamefont{J.~R.} \bibnamefont{Christiansen}},
  \bibinfo{author}{\bibfnamefont{R.}~\bibnamefont{Corke}},
  \bibinfo{author}{\bibfnamefont{N.}~\bibnamefont{Desai}},
  \bibinfo{author}{\bibfnamefont{P.}~\bibnamefont{Ilten}},
  \bibinfo{author}{\bibfnamefont{S.}~\bibnamefont{Mrenna}},
  \bibinfo{author}{\bibfnamefont{S.}~\bibnamefont{Prestel}},
  \bibinfo{author}{\bibfnamefont{C.~O.} \bibnamefont{Rasmussen}},
  \bibnamefont{and} \bibinfo{author}{\bibfnamefont{P.~Z.}
  \bibnamefont{Skands}}, \bibinfo{journal}{Comput. Phys. Commun.}
  \textbf{\bibinfo{volume}{191}}, \bibinfo{pages}{159} (\bibinfo{year}{2015}),
  \eprint{1410.3012}.

\bibitem[{\citenamefont{Marchesini et~al.}(1992)\citenamefont{Marchesini,
  Webber, Abbiendi, Knowles, Seymour, and Stanco}}]{Marchesini:1991ch}
\bibinfo{author}{\bibfnamefont{G.}~\bibnamefont{Marchesini}},
  \bibinfo{author}{\bibfnamefont{B.~R.} \bibnamefont{Webber}},
  \bibinfo{author}{\bibfnamefont{G.}~\bibnamefont{Abbiendi}},
  \bibinfo{author}{\bibfnamefont{I.~G.} \bibnamefont{Knowles}},
  \bibinfo{author}{\bibfnamefont{M.~H.} \bibnamefont{Seymour}},
  \bibnamefont{and} \bibinfo{author}{\bibfnamefont{L.}~\bibnamefont{Stanco}},
  \bibinfo{journal}{Comput. Phys. Commun.} \textbf{\bibinfo{volume}{67}},
  \bibinfo{pages}{465} (\bibinfo{year}{1992}).

\bibitem[{\citenamefont{Bellm et~al.}(2016)}]{Bellm:2015jjp}
\bibinfo{author}{\bibfnamefont{J.}~\bibnamefont{Bellm}} \bibnamefont{et~al.},
  \bibinfo{journal}{Eur. Phys. J. C} \textbf{\bibinfo{volume}{76}},
  \bibinfo{pages}{196} (\bibinfo{year}{2016}), \eprint{1512.01178}.

\bibitem[{\citenamefont{Putschke et~al.}(2019)}]{Putschke:2019yrg}
\bibinfo{author}{\bibfnamefont{J.~H.} \bibnamefont{Putschke}}
  \bibnamefont{et~al.} (\bibinfo{year}{2019}), \eprint{1903.07706}.

\bibitem[{\citenamefont{Webber}(1984)}]{Webber:1983if}
\bibinfo{author}{\bibfnamefont{B.~R.} \bibnamefont{Webber}},
  \bibinfo{journal}{Nucl. Phys. B} \textbf{\bibinfo{volume}{238}},
  \bibinfo{pages}{492} (\bibinfo{year}{1984}).

\bibitem[{\citenamefont{Andersson et~al.}(1983)\citenamefont{Andersson,
  Gustafson, Ingelman, and Sjostrand}}]{Andersson:1983ia}
\bibinfo{author}{\bibfnamefont{B.}~\bibnamefont{Andersson}},
  \bibinfo{author}{\bibfnamefont{G.}~\bibnamefont{Gustafson}},
  \bibinfo{author}{\bibfnamefont{G.}~\bibnamefont{Ingelman}}, \bibnamefont{and}
  \bibinfo{author}{\bibfnamefont{T.}~\bibnamefont{Sjostrand}},
  \bibinfo{journal}{Phys. Rept.} \textbf{\bibinfo{volume}{97}},
  \bibinfo{pages}{31} (\bibinfo{year}{1983}).

\bibitem[{\citenamefont{Das and Hwa}(1977)}]{Das:1977cp}
\bibinfo{author}{\bibfnamefont{K.~P.} \bibnamefont{Das}} \bibnamefont{and}
  \bibinfo{author}{\bibfnamefont{R.~C.} \bibnamefont{Hwa}},
  \bibinfo{journal}{Phys. Lett. B} \textbf{\bibinfo{volume}{68}},
  \bibinfo{pages}{459} (\bibinfo{year}{1977}), \bibinfo{note}{[Erratum:
  Phys.Lett.B 73, 504 (1978)]}.

\bibitem[{\citenamefont{Fries et~al.}(2003)\citenamefont{Fries, Muller, Nonaka,
  and Bass}}]{Fries:2003kq}
\bibinfo{author}{\bibfnamefont{R.~J.} \bibnamefont{Fries}},
  \bibinfo{author}{\bibfnamefont{B.}~\bibnamefont{Muller}},
  \bibinfo{author}{\bibfnamefont{C.}~\bibnamefont{Nonaka}}, \bibnamefont{and}
  \bibinfo{author}{\bibfnamefont{S.~A.} \bibnamefont{Bass}},
  \bibinfo{journal}{Phys. Rev. C} \textbf{\bibinfo{volume}{68}},
  \bibinfo{pages}{044902} (\bibinfo{year}{2003}), \eprint{nucl-th/0306027}.

\bibitem[{\citenamefont{Fries}(2004)}]{Fries:2004ej}
\bibinfo{author}{\bibfnamefont{R.~J.} \bibnamefont{Fries}},
  \bibinfo{journal}{J. Phys. G} \textbf{\bibinfo{volume}{30}},
  \bibinfo{pages}{S853} (\bibinfo{year}{2004}), \eprint{nucl-th/0403036}.

\bibitem[{\citenamefont{Greco et~al.}(2003{\natexlab{a}})\citenamefont{Greco,
  Ko, and Levai}}]{Greco:2003xt}
\bibinfo{author}{\bibfnamefont{V.}~\bibnamefont{Greco}},
  \bibinfo{author}{\bibfnamefont{C.~M.} \bibnamefont{Ko}}, \bibnamefont{and}
  \bibinfo{author}{\bibfnamefont{P.}~\bibnamefont{Levai}},
  \bibinfo{journal}{Phys. Rev. Lett.} \textbf{\bibinfo{volume}{90}},
  \bibinfo{pages}{202302} (\bibinfo{year}{2003}{\natexlab{a}}),
  \eprint{nucl-th/0301093}.

\bibitem[{\citenamefont{Greco et~al.}(2003{\natexlab{b}})\citenamefont{Greco,
  Ko, and Levai}}]{Greco:2003mm}
\bibinfo{author}{\bibfnamefont{V.}~\bibnamefont{Greco}},
  \bibinfo{author}{\bibfnamefont{C.~M.} \bibnamefont{Ko}}, \bibnamefont{and}
  \bibinfo{author}{\bibfnamefont{P.}~\bibnamefont{Levai}},
  \bibinfo{journal}{Phys. Rev. C} \textbf{\bibinfo{volume}{68}},
  \bibinfo{pages}{034904} (\bibinfo{year}{2003}{\natexlab{b}}),
  \eprint{nucl-th/0305024}.

\bibitem[{\citenamefont{Fries et~al.}(2008)\citenamefont{Fries, Greco, and
  Sorensen}}]{Fries:2008hs}
\bibinfo{author}{\bibfnamefont{R.~J.} \bibnamefont{Fries}},
  \bibinfo{author}{\bibfnamefont{V.}~\bibnamefont{Greco}}, \bibnamefont{and}
  \bibinfo{author}{\bibfnamefont{P.}~\bibnamefont{Sorensen}},
  \bibinfo{journal}{Ann. Rev. Nucl. Part. Sci.} \textbf{\bibinfo{volume}{58}},
  \bibinfo{pages}{177} (\bibinfo{year}{2008}), \eprint{0807.4939}.

\bibitem[{\citenamefont{Han et~al.}(2016)\citenamefont{Han, Fries, and
  Ko}}]{Han:2016uhh}
\bibinfo{author}{\bibfnamefont{K.~C.} \bibnamefont{Han}},
  \bibinfo{author}{\bibfnamefont{R.~J.} \bibnamefont{Fries}}, \bibnamefont{and}
  \bibinfo{author}{\bibfnamefont{C.~M.} \bibnamefont{Ko}},
  \bibinfo{journal}{Phys. Rev. C} \textbf{\bibinfo{volume}{93}},
  \bibinfo{pages}{045207} (\bibinfo{year}{2016}), \eprint{1601.00708}.

\bibitem[{\citenamefont{Majumder}(2013)}]{Majumder:2013re}
\bibinfo{author}{\bibfnamefont{A.}~\bibnamefont{Majumder}},
  \bibinfo{journal}{Phys. Rev. C} \textbf{\bibinfo{volume}{88}},
  \bibinfo{pages}{014909} (\bibinfo{year}{2013}), \eprint{1301.5323}.

\bibitem[{\citenamefont{Cao et~al.}(2016)\citenamefont{Cao, Luo, Qin, and
  Wang}}]{Cao:2016gvr}
\bibinfo{author}{\bibfnamefont{S.}~\bibnamefont{Cao}},
  \bibinfo{author}{\bibfnamefont{T.}~\bibnamefont{Luo}},
  \bibinfo{author}{\bibfnamefont{G.-Y.} \bibnamefont{Qin}}, \bibnamefont{and}
  \bibinfo{author}{\bibfnamefont{X.-N.} \bibnamefont{Wang}},
  \bibinfo{journal}{Phys. Rev. C} \textbf{\bibinfo{volume}{94}},
  \bibinfo{pages}{014909} (\bibinfo{year}{2016}), \eprint{1605.06447}.

\bibitem[{\citenamefont{Huovinen and Petersen}(2012)}]{Huovinen:2012is}
\bibinfo{author}{\bibfnamefont{P.}~\bibnamefont{Huovinen}} \bibnamefont{and}
  \bibinfo{author}{\bibfnamefont{H.}~\bibnamefont{Petersen}},
  \bibinfo{journal}{Eur. Phys. J. A} \textbf{\bibinfo{volume}{48}},
  \bibinfo{pages}{171} (\bibinfo{year}{2012}), \eprint{1206.3371}.

\bibitem[{\citenamefont{Kordell et~al.}(2022)\citenamefont{Kordell, Fries, and
  Ko}}]{Kordell:2021prk}
\bibinfo{author}{\bibfnamefont{M.}~\bibnamefont{Kordell}, \bibfnamefont{II}},
  \bibinfo{author}{\bibfnamefont{R.~J.} \bibnamefont{Fries}}, \bibnamefont{and}
  \bibinfo{author}{\bibfnamefont{C.~M.} \bibnamefont{Ko}},
  \bibinfo{journal}{Annals Phys.} \textbf{\bibinfo{volume}{443}},
  \bibinfo{pages}{168960} (\bibinfo{year}{2022}), \eprint{2112.12269}.

\bibitem[{\citenamefont{Tachibana et~al.}(2021)}]{JETSCAPE:2020uew}
\bibinfo{author}{\bibfnamefont{Y.}~\bibnamefont{Tachibana}}
  \bibnamefont{et~al.} (\bibinfo{collaboration}{JETSCAPE}),
  \bibinfo{journal}{Nucl. Phys. A} \textbf{\bibinfo{volume}{1005}},
  \bibinfo{pages}{121920} (\bibinfo{year}{2021}), \eprint{2002.12250}.

\bibitem[{\citenamefont{Buskulic et~al.}(1995)}]{ALEPH:1994cbg}
\bibinfo{author}{\bibfnamefont{D.}~\bibnamefont{Buskulic}} \bibnamefont{et~al.}
  (\bibinfo{collaboration}{ALEPH}), \bibinfo{journal}{Z. Phys. C}
  \textbf{\bibinfo{volume}{66}}, \bibinfo{pages}{355} (\bibinfo{year}{1995}).

\bibitem[{\citenamefont{Adare et~al.}(2013)}]{PHENIX:2013kod}
\bibinfo{author}{\bibfnamefont{A.}~\bibnamefont{Adare}} \bibnamefont{et~al.}
  (\bibinfo{collaboration}{PHENIX}), \bibinfo{journal}{Phys. Rev. C}
  \textbf{\bibinfo{volume}{88}}, \bibinfo{pages}{024906}
  (\bibinfo{year}{2013}), \eprint{1304.3410}.

\bibitem[{\citenamefont{Acharya et~al.}(2020)}]{ALICE:2019hno}
\bibinfo{author}{\bibfnamefont{S.}~\bibnamefont{Acharya}} \bibnamefont{et~al.}
  (\bibinfo{collaboration}{ALICE}), \bibinfo{journal}{Phys. Rev. C}
  \textbf{\bibinfo{volume}{101}}, \bibinfo{pages}{044907}
  (\bibinfo{year}{2020}), \eprint{1910.07678}.

\bibitem[{\citenamefont{Hirano and Nara}(2004)}]{Hirano:2003pw}
\bibinfo{author}{\bibfnamefont{T.}~\bibnamefont{Hirano}} \bibnamefont{and}
  \bibinfo{author}{\bibfnamefont{Y.}~\bibnamefont{Nara}},
  \bibinfo{journal}{Phys. Rev. C} \textbf{\bibinfo{volume}{69}},
  \bibinfo{pages}{034908} (\bibinfo{year}{2004}), \eprint{nucl-th/0307015}.

\bibitem[{\citenamefont{Abelev et~al.}(2008)}]{STAR:2008ftz}
\bibinfo{author}{\bibfnamefont{B.~I.} \bibnamefont{Abelev}}
  \bibnamefont{et~al.} (\bibinfo{collaboration}{STAR}), \bibinfo{journal}{Phys.
  Rev. C} \textbf{\bibinfo{volume}{77}}, \bibinfo{pages}{054901}
  (\bibinfo{year}{2008}), \eprint{0801.3466}.

\bibitem[{\citenamefont{Abelev et~al.}(2015)}]{ALICE:2014wao}
\bibinfo{author}{\bibfnamefont{B.~B.} \bibnamefont{Abelev}}
  \bibnamefont{et~al.} (\bibinfo{collaboration}{ALICE}),
  \bibinfo{journal}{JHEP} \textbf{\bibinfo{volume}{06}}, \bibinfo{pages}{190}
  (\bibinfo{year}{2015}), \eprint{1405.4632}.

\bibitem[{\citenamefont{Wang and Guo}(2001)}]{Wang:2001ifa}
\bibinfo{author}{\bibfnamefont{X.-N.} \bibnamefont{Wang}} \bibnamefont{and}
  \bibinfo{author}{\bibfnamefont{X.-f.} \bibnamefont{Guo}},
  \bibinfo{journal}{Nucl. Phys. A} \textbf{\bibinfo{volume}{696}},
  \bibinfo{pages}{788} (\bibinfo{year}{2001}), \eprint{hep-ph/0102230}.

\bibitem[{\citenamefont{Gyulassy et~al.}(2004)\citenamefont{Gyulassy, Vitev,
  Wang, and Zhang}}]{Gyulassy:2003mc}
\bibinfo{author}{\bibfnamefont{M.}~\bibnamefont{Gyulassy}},
  \bibinfo{author}{\bibfnamefont{I.}~\bibnamefont{Vitev}},
  \bibinfo{author}{\bibfnamefont{X.-N.} \bibnamefont{Wang}}, \bibnamefont{and}
  \bibinfo{author}{\bibfnamefont{B.-W.} \bibnamefont{Zhang}}, pp.
  \bibinfo{pages}{123--191} (\bibinfo{year}{2004}), \eprint{nucl-th/0302077}.

\bibitem[{\citenamefont{Salgado and Wiedemann}(2003)}]{Salgado:2003gb}
\bibinfo{author}{\bibfnamefont{C.~A.} \bibnamefont{Salgado}} \bibnamefont{and}
  \bibinfo{author}{\bibfnamefont{U.~A.} \bibnamefont{Wiedemann}},
  \bibinfo{journal}{Phys. Rev. D} \textbf{\bibinfo{volume}{68}},
  \bibinfo{pages}{014008} (\bibinfo{year}{2003}), \eprint{hep-ph/0302184}.

\bibitem[{\citenamefont{Qin et~al.}(2008)\citenamefont{Qin, Ruppert, Gale,
  Jeon, Moore, and Mustafa}}]{Qin:2007rn}
\bibinfo{author}{\bibfnamefont{G.-Y.} \bibnamefont{Qin}},
  \bibinfo{author}{\bibfnamefont{J.}~\bibnamefont{Ruppert}},
  \bibinfo{author}{\bibfnamefont{C.}~\bibnamefont{Gale}},
  \bibinfo{author}{\bibfnamefont{S.}~\bibnamefont{Jeon}},
  \bibinfo{author}{\bibfnamefont{G.~D.} \bibnamefont{Moore}}, \bibnamefont{and}
  \bibinfo{author}{\bibfnamefont{M.~G.} \bibnamefont{Mustafa}},
  \bibinfo{journal}{Phys. Rev. Lett.} \textbf{\bibinfo{volume}{100}},
  \bibinfo{pages}{072301} (\bibinfo{year}{2008}), \eprint{0710.0605}.

\bibitem[{\citenamefont{Burke et~al.}(2014)}]{JET:2013cls}
\bibinfo{author}{\bibfnamefont{K.~M.} \bibnamefont{Burke}} \bibnamefont{et~al.}
  (\bibinfo{collaboration}{JET}), \bibinfo{journal}{Phys. Rev. C}
  \textbf{\bibinfo{volume}{90}}, \bibinfo{pages}{014909}
  (\bibinfo{year}{2014}), \eprint{1312.5003}.

\bibitem[{\citenamefont{Casalderrey-Solana
  et~al.}(2014)\citenamefont{Casalderrey-Solana, Gulhan, Milhano, Pablos, and
  Rajagopal}}]{Casalderrey-Solana:2014bpa}
\bibinfo{author}{\bibfnamefont{J.}~\bibnamefont{Casalderrey-Solana}},
  \bibinfo{author}{\bibfnamefont{D.~C.} \bibnamefont{Gulhan}},
  \bibinfo{author}{\bibfnamefont{J.~G.} \bibnamefont{Milhano}},
  \bibinfo{author}{\bibfnamefont{D.}~\bibnamefont{Pablos}}, \bibnamefont{and}
  \bibinfo{author}{\bibfnamefont{K.}~\bibnamefont{Rajagopal}},
  \bibinfo{journal}{JHEP} \textbf{\bibinfo{volume}{10}}, \bibinfo{pages}{019}
  (\bibinfo{year}{2014}), \bibinfo{note}{[Erratum: JHEP09,175(2015)]},
  \eprint{1405.3864}.

\bibitem[{\citenamefont{Kumar et~al.}(2019)}]{pppaper}
\bibinfo{author}{\bibfnamefont{A.}~\bibnamefont{Kumar}} \bibnamefont{et~al.}
  (\bibinfo{collaboration}{JETSCAPE}) (\bibinfo{year}{2019}).

\bibitem[{\citenamefont{Cacciari et~al.}(2012)\citenamefont{Cacciari, Salam,
  and Soyez}}]{Cacciari:2011ma}
\bibinfo{author}{\bibfnamefont{M.}~\bibnamefont{Cacciari}},
  \bibinfo{author}{\bibfnamefont{G.~P.} \bibnamefont{Salam}}, \bibnamefont{and}
  \bibinfo{author}{\bibfnamefont{G.}~\bibnamefont{Soyez}},
  \bibinfo{journal}{Eur. Phys. J. C} \textbf{\bibinfo{volume}{72}},
  \bibinfo{pages}{1896} (\bibinfo{year}{2012}), \eprint{1111.6097}.

\bibitem[{\citenamefont{Kumar et~al.}(2023)}]{JETSCAPE:2022jer}
\bibinfo{author}{\bibfnamefont{A.}~\bibnamefont{Kumar}} \bibnamefont{et~al.}
  (\bibinfo{collaboration}{JETSCAPE}), \bibinfo{journal}{Phys. Rev. C}
  \textbf{\bibinfo{volume}{107}}, \bibinfo{pages}{034911}
  (\bibinfo{year}{2023}), \eprint{2204.01163}.

\bibitem[{\citenamefont{Fan et~al.}(2024)}]{JETSCAPE:2023ikg}
\bibinfo{author}{\bibfnamefont{W.}~\bibnamefont{Fan}} \bibnamefont{et~al.}
  (\bibinfo{collaboration}{JETSCAPE}), \bibinfo{journal}{Phys. Rev. C}
  \textbf{\bibinfo{volume}{109}}, \bibinfo{pages}{064903}
  (\bibinfo{year}{2024}), \eprint{2307.09641}.

\end{thebibliography}


\end{document}